\pdfoutput=1
\documentclass[11pt]{article}
\usepackage[preprint]{acl}
\usepackage{times}
\usepackage{latexsym}
\usepackage[T1]{fontenc}
\usepackage[utf8]{inputenc}
\usepackage{microtype}
\usepackage{inconsolata}
\usepackage{graphicx}
\usepackage{fontawesome5}

\usepackage{amsmath, amssymb, mathtools}
\usepackage{booktabs}
\usepackage{multirow}
\usepackage{url}
\usepackage{xcolor}
\usepackage{pifont}
\usepackage{xspace}

\newcommand{\benchmark}{\textsc{RepoPremium}}
\newcommand{\coreset}{\benchmark{}-\textsc{Core}}
\newcommand{\extset}{\benchmark{}-\textsc{Ext}}
\newcommand{\cellA}{\textsc{Cur-Pre}}
\newcommand{\cellB}{\textsc{Raw-Post}}
\newcommand{\cellRA}{\textsc{Raw-Pre'23}}
\newcommand{\cellRB}{\textsc{Raw-Pre'22}}
\newcommand{\cellC}{\textsc{Desc}}

\newcommand{\ndcg}[1]{\ensuremath{\text{nDCG}@#1}}

\newcommand{\fig}[1]{\cref{fig:#1}}
\newcommand{\tab}[1]{\cref{tab:#1}}
\newcommand{\secref}[1]{\cref{sec:#1}}

\usepackage[noabbrev,capitalize]{cleveref}
\crefname{equation}{equation}{equations}
\crefname{section}{Section}{Sections}
\crefname{footnote}{footnote}{footnotes}
\crefname{line}{line}{lines}
\crefname{lstlisting}{listing}{listings}
\Crefname{lstlisting}{Listing}{Listings}
\crefname{appendix}{Appendix}{Appendices}
\Crefname{appendix}{Appendix}{Appendices}

\definecolor{aigold}{RGB}{244,210, 1}
\definecolor{aigreen}{RGB}{245, 255, 249}

\definecolor{humanpurple}{RGB}{235, 222, 240}

\definecolor{commentgray}{RGB}{86, 101, 115}
\definecolor{aired}{RGB}{255,180,181}

\definecolor{findcolor}{HTML}{1D5FAE}
\definecolor{guidecolor}{HTML}{0F7F56}
\usepackage{tikz}
\usepackage{adjustbox}
\usepackage{colortbl}
\newcommand{\tagbadge}[2]{\tikz[baseline=(t.base)]\node[circle, fill=#1,
  text=white, inner sep=0pt, minimum size=1.3em,
  font=\sffamily\bfseries\fontsize{6.5}{7}\selectfont] (t) {#2};}
\newcommand{\fone}{\texorpdfstring{\tagbadge{findcolor}{F1}\xspace}{F1}}
\newcommand{\ftwo}{\texorpdfstring{\tagbadge{findcolor}{F2}\xspace}{F2}}
\newcommand{\fthree}{\texorpdfstring{\tagbadge{findcolor}{F3}\xspace}{F3}}
\newcommand{\gone}{\texorpdfstring{\tagbadge{guidecolor}{G1}\xspace}{G1}}
\newcommand{\gtwo}{\texorpdfstring{\tagbadge{guidecolor}{G2}\xspace}{G2}}
\newcommand{\gthree}{\texorpdfstring{\tagbadge{guidecolor}{G3}\xspace}{G3}}

\usepackage{listings}
\usepackage{parcolumns}
\newcommand{\codefont}{\fontfamily{lmtt}\selectfont}
\lstdefinestyle{datalogstyle}{
	basicstyle={\codefont\small},
	xleftmargin={6pt},
	xrightmargin={6pt},
	breakindent=0pt,
	frame=tb,
	stepnumber=1,
	firstnumber=1,
	numberfirstline=true,
	tabsize=2,
	showtabs=false,
	showspaces=false,
	showstringspaces=false,
	extendedchars=true,
	breaklines=true,
	columns=fullflexible,
	keepspaces=true,
	escapeinside={@}{@},
	firstnumber=last,
	captionpos=b,
	commentstyle=\color{black!65},
	numberstyle=\tiny\color{black!65},
	breakatwhitespace=false,
	numbersep=5pt,
	aboveskip={0.8\baselineskip},
	belowskip={0.2\baselineskip},
	backgroundcolor=\color{aigreen},
}
\title{Embeddings vs.\ Grep: Rethinking the Dense Advantage in Code Search}

\author{Siqiao Xue\thanks{Work done at Ant Group.} \quad Ziyin Zhang \quad Zihan Liao \quad Jin Qin \quad Yixiang Mu\footnotemark[1] \quad Hang Yu \\[4pt]
daenerystargaryen@sjtu.edu.cn \\ [4pt]
  Ant Group \\[4pt]
  \faGithub\ \small \url{https://github.com/hq-bench/repo-premium} \\
  \raisebox{-0.15em}{\includegraphics[height=1em]{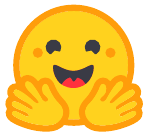}}\ \small \url{https://huggingface.co/datasets/hq-bench/repo-premium}}

\begin{document}
\maketitle

\begin{abstract}
Code-search benchmarks show dense encoders well ahead of BM25, yet
coding agents do well with \texttt{grep}. What does this dense
advantage measure? Existing benchmarks cannot tell: they use curated
queries, old code and a handicapped BM25. To find out, we build
\benchmark{}, a realistic repository-level code-search benchmark. It
ranks files in real repositories, with curated and raw queries in
Python, JavaScript, TypeScript, Go, Rust and Java, pull requests from 2022 to
2026, and a preregistered test pool over $91$ repositories. Our
experiments trace the advantage to two sources: evaluation choices
that favor dense retrieval, such as splitting files into chunks for
BM25 and counting edited tests as relevant, and queries that lack the
code's rare identifiers. Expanding the query with LLM-written identifiers
(generate-then-grep) helps BM25 far more than dense retrieval, letting it match dense
retrieval in every Python pool, even on pull requests opened after the
LLM was released. No method we tried predicts which retriever
will win on a given query. From these results we derive three
guidelines: measure dense gains against a fair lexical baseline, fuse
dense retrieval with BM25, and add the code's words to queries that
lack them.
\end{abstract}

\section{Introduction}
\label{sec:intro}

Coding agents are tested on ever newer repository tasks
\citep{zhang2025swebenchlive, huang2026deepswe}, but an end-to-end score
does not say whether search, reasoning or editing failed. Retrieval
benchmarks test search alone. On them, code-specialized encoders beat
BM25 by wide \ndcg{10} margins \citep{guenther2024jinacode,
li2024coir}. Yet one practitioner report found that embeddings were not
the bottleneck for a SWE-bench agent that searched with \texttt{grep}
and \texttt{find} \citep{augment2025}. Teams build embedding indexes
because of benchmark gaps, so the conflict matters. What does the
dense advantage buy, and where does it come from?

Existing evidence cannot settle this. Embedding benchmarks mostly pair
curated or docstring-style queries with old code
\citep{husain2019codesearchnet, muennighoff2023mteb}; agent reports use
raw issue or pull-request (PR) text \citep{jimenez2024swebench}.
Lexical baselines often reuse the dense retriever's chunking. Weak
baselines are a known source of false gains in neural ranking
\citep{lin2019neuralhype, yang2019critically, kamphuis2020bm25}, but
their effect in repository code search has not been measured.
Gold labels also differ. SWE-bench's gold patches contain no test files
\citep{jimenez2024swebench}. A PR's list of changed files does, and
those tests count as relevant unless they are removed. Our own first
version kept them. Tests can be fair targets elsewhere, so we
report both label rules. Finally, benchmark code predates the encoders'
training cutoffs, so the advantage could also reflect contamination
\citep{magar2022contamination, sainz2023nlpevaltrouble}.

To answer this, we build \benchmark{}, a realistic repository-level
code-search benchmark over real repositories with file-level labels. \textbf{Repository localization pipelines rank
files first, and file labels come straight from the merged change}
(\cref{sec:benchmark}). Within each repository we fix the code
snapshot, the label rule and the protocol, and vary only where the query
comes from and when the change was made. The core has $1{,}265$ queries
over ten SWE-bench Verified repositories: curated problem statements
\citep{openai2024swebenchverified} and raw GitHub PR text from 2022,
2023 and 2026.\footnote{Each raw pool takes PRs merged in one window:
2022 (\cellRB{}), 2023 (\cellRA{}) and on or after 2026-01-01
(\cellB{}). We skip 2024--2025 so that every window lies clearly before
or after the encoders' training cutoffs; see Step~2 of
\cref{sec:pipeline}, \tab{cells} and \cref{sec:construction}.} Each
main analysis is thus repeated across query sources and years. An extension
set adds five languages, nine held-out repositories and a
$91$-repository preregistered test pool (\tab{cells}). The two
preregistered tests are scored on the labels as released (edited tests
included), with source-file results alongside. We use ``grep'' for
lexical search and measure it with ranked BM25, so generate-then-grep
is BM25 on an expanded query; only the agent evaluation
(\cref{sec:agentsummary}) lets agents call \texttt{grep} directly on
$96$~tasks.

We call dense retrieval's \ndcg{10} lead over BM25 the \emph{dense
premium} and ask three questions. How large is it under one consistent
protocol? Which queries carry it? Should a system route each query to a
retriever or fix one setup? Hybrid retrieval is known to help
\citep{thakur2021beir}, and LLM query expansion helps both BM25 and
dense retrieval \citep{wang2023query2doc, gao2023hyde}. We use them as
controlled interventions,
\textbf{asking which retriever expansion helps and why}, and test our
advice on preregistered held-out repositories.

\begin{figure}[t]
\centering
\includegraphics[width=\columnwidth]{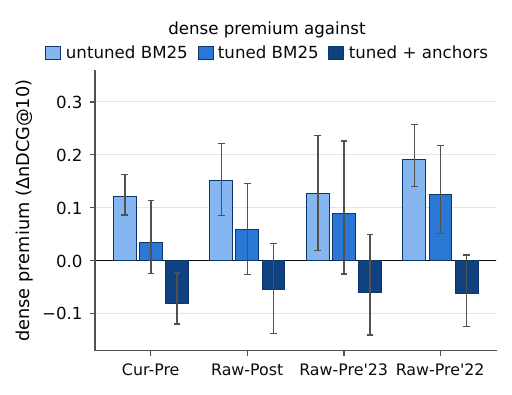}
\caption{Dense retrieval's advantage over BM25 (\ndcg{10}, canonical
encoder \texttt{jina-code-embeddings-0.5b}, source-file labels) in the
four Python pools, against three
versions of BM25: untuned (the dense retriever's chunking, $c=1{,}000$),
tuned (whole files), and tuned with LLM-written anchors added to the
query (generate-then-grep). Tuning BM25 shrinks the premium to
$+0.03$--$+0.13$; adding anchors removes it in every pool, with BM25
significantly ahead on the curated pool. Means with repository-clustered 95\% CIs
(\cref{tab:genarm-tuned}).}
\label{fig:overview}
\end{figure}

We evaluate thirteen dense encoders, plus SPLADE
\citep{formal2021splade} and ColBERTv2 \citep{santhanam2022colbertv2}.
Our experiments identify where the premium comes
from:

\noindent\fone \textbf{Evaluation choices inflate the dense advantage;
a smaller lead survives on raw queries.} Indexing whole files for BM25
cuts the curated premium from $+0.121$ to $+0.034$; counting edited
tests as relevant doubles the \cellB{} premium (\cref{fig:overview},
\cref{sec:f0}). Against whole-file BM25 the encoder still leads the raw
pools, mostly through Recall@10; on curated queries its interval covers
zero.

\noindent\ftwo \textbf{Adding the code's words closes what remains.}
The premium sits on queries that lack the code's rare words, and
LLM-written identifiers help BM25 far more than the encoder. From the same
expanded query the encoder gains a fraction of what BM25 gains, and
prose stripped of identifiers changes nothing. BM25 then matches or beats dense retrieval in
every Python pool, even on PRs newer than the LLM. The LLM may know the
repository, so this replaces the premium but does not explain it
(\cref{sec:tau,sec:genarm}).

\noindent\fthree \textbf{The two retrievers win different queries, and
nothing we tried predicts which.} Classifiers, LLM judges and a
preregistered router all fail. A fixed fusion of the two beats either
retriever alone and held on $539$ preregistered lockbox queries
(\cref{sec:predictable,sec:routing}).

From these findings we derive three guidelines for code search:

\noindent\gone \textbf{Measure dense gains against a fair lexical
baseline.} Index whole
files for BM25, strip templated PR text, and state which edited files
count as relevant before comparing against dense retrieval.

\noindent\gtwo \textbf{Keep dense retrieval and fuse it with BM25.} A
fixed rank fusion beats dense alone on development data and on $539$
lockbox queries, and no router beat the fusion. In agents, the LLM
matters more than the search tool (\cref{sec:agentsummary}).

\noindent\gthree \textbf{When a query names no code, add the code's
words before searching.} One LLM call that drafts the missing
identifiers lets whole-file BM25 match an embedding index. The call
recurs per query, so it suits repositories searched rarely
(\cref{app:genarm-moved}).

\section{Related Work}
\label{sec:related}

Agent suites such as SWE-bench \citep{jimenez2024swebench} and DeepSWE
\citep{huang2026deepswe} run the whole agent loop. Code-retrieval
benchmarks such as CoIR \citep{li2024coir} pair docstring- or web-style
queries with old code. None of them varies query mining and
training-cutoff date together. Their labels also differ:
SWE-bench gold patches contain no tests, while a PR's file list keeps
every edited file. It is well known that lexical and dense retrievers
fail on different queries and that fusing them helps
\citep{cormack2009rrf, thakur2021beir}. Our $\tau$ is the highest
inverse document frequency (IDF) among tokens shared by the query and
the first source file the change edited. We compute IDF over the
repository's chunks. It adapts the MaxIDF predictor
\citep{hauff2008preretrieval} but needs the gold file; its proxy
$\widehat\tau$ uses the query alone. LLM query expansion helps BM25
\citep{wang2023query2doc} and, through generated pseudo-documents, dense
retrieval \citep{gao2023hyde}. We use it as a controlled intervention,
not as a proposed method.
The closest work, SWE-Explore \citep{zhang2026sweexplore}, also isolates
retrieval: it scores the code regions that successful agents read,
under a line budget, and its dense baseline uses static word
embeddings. We score the files the change edited, use current code
encoders, and vary the lexical baseline, the label rule and the query
source. \cref{app:related-ext} gives further references.

\section{\benchmark{} Construction}
\label{sec:benchmark}

\subsection{Task and Data}
\label{sec:task-data}

\paragraph{Task.} \benchmark{} is file retrieval inside one repository.
\emph{Input:} a natural-language description of a code change (an issue
statement or a PR's title and body) and a pinned snapshot of its
repository. \emph{Output:} a ranking of the snapshot's files.
\emph{Relevant:} the non-test source files the described change
edited. Test files stay in the ranking but count as non-relevant, and
every retriever ranks the same files for the same query (an example in
\cref{fig:example-queries}).

\paragraph{Two sets: \coreset{} and \extset{}.} The data has two sets
(\tab{cells}). \coreset{} carries the main results. Its four query
pools differ in \emph{how} the query was written and \emph{when} the
change was made, but share ten Python repositories, one code snapshot
per repository (except \cellRA{}, Step~6) and one label rule. The pools
are curated statements from before the encoders' cutoffs (\cellA{}) and
raw PR text from three windows: after the cutoffs (\cellB{}), 2023
(\cellRA{}) and 2022 (\cellRB{}). Because the code is shared, differences between pools
come from the queries. The pools still describe different changes, so
these comparisons are descriptive. The core is Python-only because
SWE-bench Verified, the source of the curated queries, is Python-only,
and the raw pools must use the same repositories.

\extset{} (extension) checks whether the core results hold elsewhere.
Each pool uses its own repositories. The set has raw PRs in JavaScript,
TypeScript, Go and Rust; curated Multi-SWE-bench statements outside
Python; nine held-out Python repositories; and DeepSWE tasks written by
benchmark authors. A matched replication also presents each of $360$
changes twice, as the raw PR and as an LLM rewrite, so only the wording
changes.
All pools use the same label rule, and their raw PR text goes through
the same filters and stripper. DeepSWE and the TypeScript, Go and Rust
PRs form the preregistered lockbox, on which we tested routing and
fusion once.

\paragraph{Scale.} \coreset{} holds $1{,}265$ queries over $10{,}200$
indexed files (from $35$ to $2{,}752$ per repository;
\tab{repostats}). \extset{} adds $1{,}560$ queries, counting
both views of each matched task, from five more languages
(JavaScript, TypeScript, Go, Rust and Java), nine
held-out Python repositories and $88$ DeepSWE repositories
(\tab{cells}).

\begin{figure}[t]
\centering
\footnotesize
\raggedright
{\color{findcolor}\textbf{One change, two wordings}}\quad \texttt{pandas}
PR 58032\\[2pt]
\colorbox{humanpurple}{\strut\textbf{Gold}\ \ \texttt{pandas/core/frame.py}}

\vspace{3pt}
\textbf{Curated} (LLM rewrite of the change)
\begin{lstlisting}[basicstyle=\codefont\footnotesize,aboveskip=2pt,belowskip=2pt]
DataFrame.reset_index does not accept a tuple for the 'names' parameter, despite documentation stating it should.
\end{lstlisting}
{\scriptsize BM25: gold at rank \textbf{1}\qquad dense: rank 7}

\vspace{4pt}
\textbf{Raw} (PR title; the body is template text, stripped)
\begin{lstlisting}[basicstyle=\codefont\footnotesize,aboveskip=2pt,belowskip=2pt]
@\colorbox{aired}{\strut DOC:}@ DataFrame.reset_index names param can't be a tuple as @\colorbox{aired}{\strut docs}@ state
\end{lstlisting}
{\scriptsize BM25: rank 8, top hit \texttt{doc/source/conf.py}\qquad dense:
gold at rank \textbf{1}}
\caption{The same change worded two ways, from the matched replication
(\texttt{jina-code-0.5b} against BM25 at $c=1{,}000$). The curated
rewrite states the behavior in the code's terms, and BM25 ranks the gold
file first. In the raw PR title, the words \emph{DOC} and \emph{docs}
(highlighted) pull BM25 toward documentation files, while dense
retrieval still ranks the gold file first. Gold sets are the source
files the change edited (Step~3).}
\label{fig:example-queries}
\end{figure}

\begin{figure*}[t]
\centering
\includegraphics[width=\linewidth]{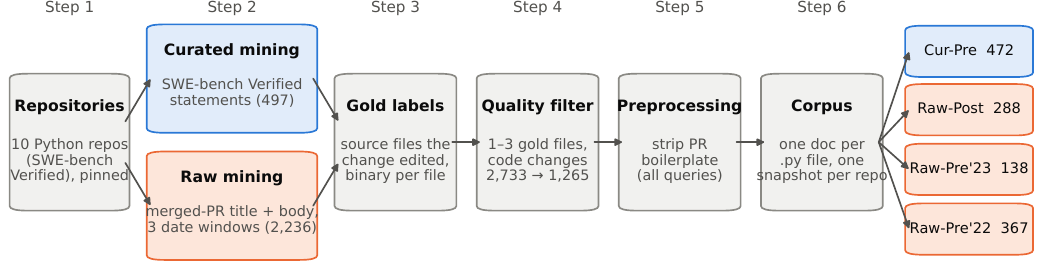}
\caption{Construction pipeline of the four \coreset{} pools (Steps 1--6 of
\cref{sec:pipeline}); the \extset{} pools of \tab{cells} reuse Steps
3--5 on their own repositories. Curated SWE-bench Verified statements (blue) and
raw merged-PR text (orange) share the label rules, preprocessing and
one code snapshot per repository (\cellRA{} uses a 2024-Q1 snapshot).
Text filters apply to raw PRs only. Of $2{,}733$ candidates, $1{,}265$
remain in the four core pools.}
\label{fig:construction}
\end{figure*}

\begin{table*}[t]
  \centering
  \small
  \setlength{\tabcolsep}{6pt}
  \renewcommand{\arraystretch}{1.12}
  \begin{adjustbox}{max width=\textwidth}
  \begin{sc}
  \begin{tabular}{l l l l r r r}
    \toprule
    Pool & Query text & Coverage date & Languages & Repos & Files & Queries \\
    \midrule
    \rowcolor{findcolor!10}
    \textbf{\coreset{}} & & & & & & \\
    \cellA{}  & SWE-bench issue & 2013--23 & Python & 10 & $10{,}200$ & 472 \\
    \cellB{}  & PR title + body & 2026 & Python & 9 & $9{,}619$ & 288 \\
    \cellRA{} & PR title + body & 2023 & Python & 10 & $10{,}469$ & 138 \\
    \cellRB{} & PR title + body & 2022 & Python & 10 & $10{,}200$ & 367 \\
    \multicolumn{6}{r}{\emph{total}} & $1{,}265$ \\
    \rowcolor{guidecolor!10}
    \textbf{\extset{}} & & & & & & \\
    Cross-language & PR title + body & 2022--23 & JS, TS, Go, Rust & 4 & $37{,}712$ & 458 \\
    Multi-SWE-bench & curated issue & -- & TS/JS, Rust & 3 & $26{,}000$ & 134 \\
    Matched repl. & PR + LLM rewrite & both sides & 5 languages & 36 & $154{,}041$ & $360{\times}2$ \\
    Held-out Python & PR title + body & $\geq$2025-04 & Python & 9 & $3{,}702$ & 139 \\
    DeepSWE & task spec & -- & 5 languages & 88 & $31{,}288$ & 109 \\
    \multicolumn{6}{r}{\emph{total}} & $1{,}560$ \\
    \bottomrule
  \end{tabular}
  \end{sc}
  \end{adjustbox}
  \caption{Query pools. \coreset{} carries the main results over one
  shared corpus; \extset{} tests whether they hold elsewhere, each pool
  on its own repositories. \emph{Coverage date}: the period the
  pool's queries come from (PR merge dates; for \cellA{}, SWE-bench
  issue dates); the
  encoders' training cutoffs fall in 2024--25, so \coreset{} pools lie
  clearly before (\cellA{}, \cellRA{}, \cellRB{}) or after (\cellB{}).
  Benchmark pools (--) keep their released tasks; the matched
  replication samples both sides of the cutoff. \emph{Files}: indexed
  files summed over a pool's repositories (\cellRA{} uses a 2024-Q1
  snapshot; the matched replication takes each repository's median
  over its tasks' base commits). \emph{Queries}: queries that pass the filters; the matched
  replication scores $351$ of its $360$ tasks (\cref{sec:scaleup}).
  \emph{5 languages}: matched replication, Python, Java, Go, Rust and
  JS/TS repositories sampled as one stratum; DeepSWE, Python,
  JavaScript, TypeScript, Go and Rust.
  Routing and fusion were developed on the $472$ \cellA{} and $132$
  original \cellB{} queries ($604$) and tested once on a preregistered lockbox:
  DeepSWE and the TypeScript, Go and Rust PRs ($539$ queries).}
  \label{tab:cells}
\end{table*}

\paragraph{Why rank whole files, not functions or lines?}
Repository-level work starts by choosing which files to edit: an
engineer or an agent opens a few files and edits within them.
Localization pipelines rank files before functions or lines
\citep{xia2024agentless, chen2025locagent}. SWE-bench's retrieval
baseline ranks whole files with BM25 \citep{jimenez2024swebench}. Bug
localization has long ranked files against bug reports
\citep{zhou2012buglocator}. Finer-grained work ranks functions or line
regions \citep{reddy2026swerank, zhang2026sweexplore}. File labels also
come straight from the merged change, so no annotator judges relevance.
Scoring is still fine-grained: the encoder embeds $1{,}000$-character
chunks, and a file takes its best chunk's score. The task is not easy
either: repositories hold up to $2{,}752$ files, and no pipeline we ran
exceeds $0.7$ \ndcg{10}.

\subsection{Construction Pipeline}
\label{sec:pipeline}

Six steps build the \coreset{} pools (\cref{fig:construction}). The
\extset{} pools reuse the label rule and, for raw PR text, the filters
and stripper (Steps 3--5), but index their own repositories; the
appendix details each pool and its results.

\paragraph{Step 1: Repositories.} The core uses ten of the twelve
SWE-bench Verified repositories; the other two have too few merged
PRs. \cellB{} covers nine, as \texttt{sphinx} merged too few in 2026.

\paragraph{Step 2: Queries.} \cellA{} takes each SWE-bench Verified
issue statement. The raw pools take the title and body of merged PRs
from each date window, most recent first (up to $100$ per repository,
$40$ for \cellRA{}).

\paragraph{Step 3: Gold labels.} The gold set is the indexed source
files the change edited, each with binary relevance.
\cellA{} reads them from the SWE-bench gold patch, which has no tests.
For raw PRs we drop edited tests and unindexed files (documentation,
changelogs) from the changed-file list, matching SWE-bench's oracle
files. Because this choice moves the premium, we also score with tests
included (\cref{sec:test-gold}).

\paragraph{Step 4: Quality filter.} A query is kept if it has one to
three gold files and every one exists at the snapshot; a change that
creates its source file is dropped rather than relabeled. Raw PRs must
also describe a code change (at least $100$ characters, at most $20$
edited files, not docs-, CI- or test-only). Of $2{,}733$ candidates, $1{,}265$ remain (\tab{funnel}):
changes that edit a few existing source files.

\paragraph{Step 5: Preprocessing.} A regex stripper removes PR templates,
checklists and HTML comments that otherwise swamp BM25
(\cref{sec:f5,app:stripper}).

\paragraph{Step 6: Corpus.} Each repository is pinned at the base commit
of its most recent SWE-bench Verified instance, and every \texttt{.py}
file is one document; \cellRA{} uses a 2024-Q1 snapshot. \cellB{}'s
queries and changes therefore postdate the encoders, but the indexed
code does not. Re-scoring each task at its own base commit moves no
tuned premium by more than $0.025$ (\cref{app:percommit}). Further
construction details are in \cref{app:benchmark-details}.

\paragraph{Release.} We release the queries, file-level labels with a
test-file flag, pinned commits, the stripper and all evaluation
harnesses (\cref{app:release}).

\section{Experimental Setup}
\label{sec:methods}

\paragraph{Retrievers.} We compare BM25 (BM25Okapi over chunks), a
literal IDF-weighted \texttt{grep} baseline (reported separately as
``grep'' in appendix tables), and thirteen \texttt{sentence-transformers}
encoder checkpoints. The encoders come from six families:
\texttt{jina-code-embeddings}, \texttt{Qwen3-Embedding},
\texttt{C2LLM}, \texttt{F2LLM}, \texttt{E5} and \texttt{gte-Qwen2}. All
run without instruction prefixes. Unless noted, the canonical encoder
is \texttt{jina-code-embeddings-0.5b} (\cref{app:methods-details}).

\paragraph{Chunking and the lexical baseline.} Documents are chunked at
$c=1{,}000$ characters with a 10-line overlap, and chunk scores are
max-pooled to a file score. Chunk size matters: dense encoders truncate
long inputs, while BM25 prefers large chunks. We use $c=1{,}000$ as the
common protocol. The tuned lexical baseline is BM25 over whole files,
which is BM25's best setting on the raw pools and within $0.002$ of its
best ($c=10{,}000$) on \cellA{} (\cref{sec:f0}).

\paragraph{Metric.} The per-query premium is
$\Delta(q) = \ndcg{10}_\text{dense}(q) - \ndcg{10}_\text{BM25}(q)$,
scored per file with binary relevance against source-file gold sets.
We average it over queries and also report Recall@10 and MRR@10.

\paragraph{Query distinctiveness.} For a stripped query with token set
$q$ and the token set $g$ of its first gold file (the first indexed,
non-test file in the change's list),
\begin{equation}
\tau(q) \;=\; \max_{t \in q \cap g}\, \text{IDF}(t),
\label{eq:tau}
\end{equation}
with IDF over the repository's chunk corpus: the MaxIDF predictor of
\citet{hauff2008preretrieval} restricted to shared tokens. It needs the
gold file, so it is an offline statistic. A deployable proxy $\widehat\tau$ is
the mean IDF of the stripped query's five rarest tokens, computed from
the query alone.

\paragraph{Statistics.}
\label{sec:stats-method}
Intervals are $95\%$ bootstraps, resampling repositories
for comparisons between pools. Where we say an intervention does not
raise the premium, its $90\%$ interval lies below a preset $+0.05$
\ndcg{10} margin (\cref{app:methods-details}).

\paragraph{Query sets.} Pool-level results use the full raw pools of
\tab{cells}. Analyses that needed further retrieval runs use $461$ raw
queries chosen before we unified the gold rule. These are the
thirteen-encoder sweep, HyDE, learned sparse and late-interaction
retrievers, the reranker, anchor ablation, per-commit snapshots, the
label audit, the $\tau$ analyses and the robustness battery. On this
subset every canonical ranking is identical (\cref{sec:construction}). Routing and
fusion keep the development pool they were configured on.

\section{Results and Analysis}
\label{sec:findings}
\label{sec:analysis}

\subsection{How Large Is the Premium?}
\label{sec:f0}
\paragraph{Chunking decides the sign; whole-file BM25 removes most of
the gap.} On \texttt{django}, dense retrieval over whole files loses to
BM25 ($\Delta=-0.153$) because the encoders truncate long files;
chunking at $c=1{,}000$ flips this to $+0.119$. BM25, in contrast, gains
from larger windows. Against whole-file BM25 the premium falls to $+0.034$ on
\cellA{} and $+0.059$ to $+0.125$ on the raw pools
(\cref{tab:genarm-tuned}). No encoder
keeps a significant premium on \cellA{} or \cellB{}, and seven of
thirteen fall significantly below zero on \cellA{}
(\cref{tab:chunk-rebase}). BM25's setting was picked on the evaluation
queries, so these are lower bounds for dense retrieval.

\paragraph{Counting test files as relevant inflates the premium.} Dense
retrievers rank a test file first far more often than BM25 ($59\%$
against $34\%$ of curated queries). Adding edited tests to the gold
sets therefore raises the tuned premium from $+0.034$ to $+0.165$ on
\cellA{} and from $+0.059$ to $+0.136$ on \cellB{}
(\cref{sec:test-gold}). With either rule applied to every pool, the
curated and raw pools are indistinguishable. Mixing the rules,
as happens when SWE-bench gold patches are scored next to PR file
lists, widens the \cellB{}-minus-\cellA{} gap from $+0.025$ to
$+0.102$.

\paragraph{What survives is mostly coverage.} Dense retrieval
places more gold files in its top 10 (Recall@10 $+0.10$ to $+0.19$) but
ranks its first hit significantly higher only on \cellRB{}
(\cref{tab:metrics}). Outside Python, a modest premium remains on
TypeScript and Rust but not on Go (single repositories;
\cref{app:idsplit}).

Much of the reported premium thus comes from the setup. Against
whole-file BM25 the curated premium nearly vanishes
($+0.121\to+0.034$), and the raw premiums shrink by a third to over a
half. Counting test files as relevant inflates it again. What survives
is mostly wider coverage.

\subsection{Supplying the Missing Evidence}
\label{sec:genarm}
\noindent If the premium reflects lexical evidence a query lacks,
supplying that evidence should help BM25 more than dense retrieval.

\paragraph{Generate-then-grep catches up with dense retrieval.} A
generator (\texttt{qwen35-122B}, temperature $0$, one frozen prompt)
reads the stripped query, but never gold paths or patches. It writes
identifiers, file and module hints, APIs, error terms and a code
skeleton, and we prepend this text to the query for BM25. With BM25
over whole files in both arms, generation adds $0.11$--$0.19$ \ndcg{10}
to BM25, and generate-then-grep
is never behind the canonical encoder. It leads significantly on
\cellA{}, by $0.081$. Its lead on the raw pools is $0.054$--$0.062$, but
those intervals cover zero, so a small dense lead remains possible
(\tab{genarm-tuned}). Whole files are not needed: with BM25 at the
encoder's $1{,}000$-character chunks, the premium after expansion is
$-0.052$ to $+0.012$.

\begin{table}[t]
\centering
\small
\setlength{\tabcolsep}{3.5pt}
\begin{sc}
\begin{adjustbox}{max width=\columnwidth}
\begin{tabular}{lrrrrr}
\toprule
& & \multicolumn{2}{c}{\textbf{BM25, whole files}} & & \\
\cmidrule(lr){3-4}
\textbf{Pool} & \textbf{dense} & \textbf{query} & \textbf{+anch.} & \textbf{prem.} & \textbf{after} \\
\midrule
\cellA{}  & 0.532 & 0.498 & 0.613 & $+0.034$ & $\mathbf{-0.081}$ \\
\cellB{}  & 0.597 & 0.539 & 0.651 & $+0.059$ & $-0.054$ \\
\cellRA{} & 0.544 & 0.455 & 0.603 & $+0.089$ & $-0.059$ \\
\cellRB{} & 0.572 & 0.447 & 0.635 & $+0.125$ & $-0.062$ \\
\bottomrule
\end{tabular}
\end{adjustbox}
\end{sc}
\caption{Generation against a tuned lexical baseline (\ndcg{10},
source-file labels). Both BM25 arms score whole files; dense is the
canonical encoder at $c=1{,}000$. ``Premium'' is dense minus BM25 on the
original query, ``after'' dense minus BM25 on the query plus generated
anchors; $n=472$, $288$, $138$, $367$. Bold: the repository-clustered
95\% interval excludes zero. Generate-then-grep is never behind
dense retrieval in point estimate and is significantly ahead on the
curated pool.}
\label{tab:genarm-tuned}
\end{table}

\paragraph{The added words help BM25 far more than the encoder.} When
the encoder embeds the expanded query, it gains $+0.03$ to $+0.07$,
against $+0.12$ to $+0.21$ for BM25; the difference excludes zero in
every pool (\tab{gen-controls}). Identifiers, APIs and error terms
alone close the premium; prose search phrases do not
(\cref{app:anchor-split}). Pseudo-relevance feedback (RM3, default
setting; \citealp{lavrenko2001rm}) even lowers BM25.

\paragraph{The result holds after the generator's release and against
other dense arms.} On held-out PRs opened after the generator's release,
generate-then-grep still beats dense retrieval by $0.094$
(\cref{app:genarm-moved}). Memorized changes do not explain this,
though knowledge of the repositories might.
\texttt{Qwen3-Embedding-8B} keeps a larger premium on two of the three
raw pools, but is again never ahead after generation. SPLADE
\citep{formal2021splade} and ColBERTv2 \citep{santhanam2022colbertv2}
stay below the canonical encoder (\tab{gen-controls}, \tab{encoder2}).
Smaller and weight-matched generators close two thirds to $80\%$ of the
premium, so the direction holds but the size depends on the generator.

When an LLM writes the missing identifiers into the query, BM25 is
never behind dense retrieval in any pool (point estimates), and the
same text helps the encoder far less. Generated identifiers can replace
the premium. Since the generator may know the repository, this does not
show where the premium comes from.

\subsection{Fuse Once Instead of Routing Per Query}
\label{sec:budget}
\label{sec:routing}
\paragraph{Fusion beats either retriever and a reranker.} Reciprocal
rank fusion of dense retrieval and whole-file BM25 raises \ndcg{10} by
$+0.060$ over dense alone on the development pool; fusing with
generate-then-grep instead, at one LLM call per query, raises it by
$+0.089$. A listwise reranker (\texttt{jina-reranker-v3}) over the same
candidates does no better than fusion.

\paragraph{A preregistered router misses its criterion; the fixed
fusions hold.} A router picks dense retrieval, BM25 or their fusion for
each query, and the oracle headroom is large. We built about $30$
candidate routers on query, list-agreement and score features. Only one
had a development gain that excluded zero. We froze it and scored it
once on a lockbox that no earlier analysis had touched ($539$ queries,
$91$ repositories). It gained $+0.005$ over always using the fusion,
against $+0.077$ for the oracle, and its interval covered zero, so it
missed its preregistered criterion. Both fixed fusions were configured
on development data only. On the lockbox they beat dense retrieval by
$+0.019$ and $+0.045$ on the labels as released, and by $+0.027$ and
$+0.062$ with source-file labels. They still beat it when repositories
are weighted equally (\cref{app:decision}).

A fixed fusion is therefore the robust choice: it beats either retriever
alone and held up on the lockbox, and per-query routing did not
detectably beat it (\cref{tab:budget}).

\begin{table}[t]
\centering
\small
\setlength{\tabcolsep}{4pt}
\begin{sc}
\begin{adjustbox}{max width=\columnwidth}
\begin{tabular}{llrr}
\toprule
\textbf{Pipeline} & \textbf{Per query} & \textbf{Dev} & \textbf{Lockbox} \\
\midrule
BM25, whole files & none & 0.507 & 0.432 \\
Dense & embed & 0.550 & 0.533 \\
RRF(dense, BM25) & embed & 0.609 & 0.551 \\
Reranker & +rerank & 0.574 & --- \\
Router (frozen) & embed & --- & 0.557 \\
Generate-then-grep & LLM & 0.625 & 0.534 \\
RRF(dense, gen.) & LLM+embed & \textbf{0.638} & \textbf{0.578} \\
\bottomrule
\end{tabular}
\end{adjustbox}
\end{sc}
\caption{Fixed pipelines by per-query cost (\ndcg{10}). Dev: the
development pool D (\cellA{} and the original \cellB{} queries; $604$
queries, $10$ repositories), source-file labels;
lockbox: DeepSWE and the TypeScript, Go and Rust pools ($539$ queries,
$91$ repositories), scored once as preregistered on the labels as
released. Dense is \texttt{jina-code-0.5b}; the reranker reorders the
union of the dense and BM25 lists; the router chooses per query among
dense, BM25 and their RRF. \emph{Per query}: work beyond BM25 (embed:
one query embedding; rerank: \texttt{jina-reranker-v3}; LLM: one generation
call). ---: not run (the router is fit on dev; the reranker was not
preregistered). Bold: highest score. BM25's chunk size was chosen on
dev. Differences quoted in the text are computed before rounding.}
\label{tab:budget}
\end{table}

\subsection{Which Queries Carry the Premium}
\label{sec:tau}
\begin{figure}[t]
\centering
\includegraphics[width=\columnwidth]{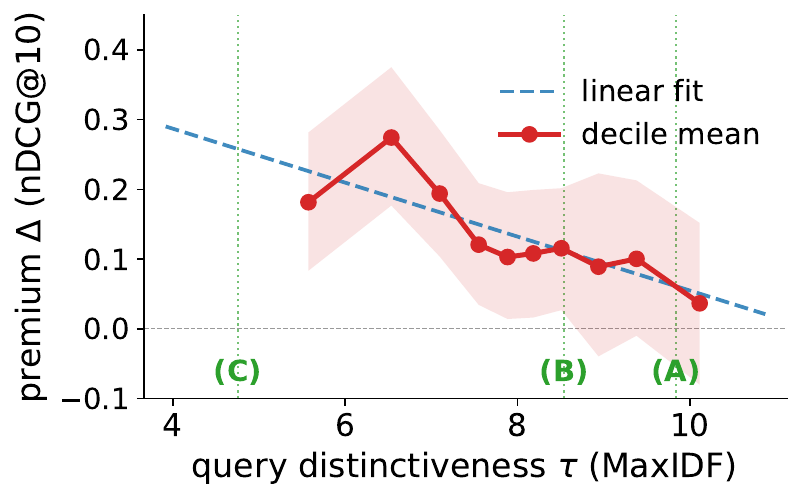}
\caption{The dense-over-BM25 premium $\Delta$ decays with query
distinctiveness $\tau$ (dense at $c=1{,}000$ against BM25 at
$c=1{,}000$; source-file labels). Decile means with query-bootstrap 95\%
CIs over the $n{=}604$ curated and original raw queries; dashed: linear
fit. Quartile means fall monotonically; deciles show the noise around
that trend. Dotted verticals mark the exemplar queries of
\cref{app:intro_exemplars}.}
\label{fig:decay}
\end{figure}

\paragraph{The premium shrinks as the shared tokens get rarer.} On
the $604$ \cellA{} and original \cellB{} queries, the premium falls from
$+0.217$ in the lowest $\tau$ quartile to $+0.050$ in the highest
(\cref{fig:decay}). Against whole-file BM25, it falls from $+0.125$ to
$-0.046$. Per query the link is weak ($r=-0.124$), so the claim is about group
means. The slope is negative in 37 of 38 slices (mining regimes,
thirteen encoders, five languages), but we chose $\tau$ on those
slices. A preregistered test on $29$ held-out repositories had the same
sign, with an interval covering zero (\cref{app:law-slices}).

\paragraph{Removing anchors hurts both retrievers.} We replace the five
rarest tokens a query shares with its first gold file by synthetic
strings, and compare against an IDF-matched placebo. If dense retrieval
did not rely on these tokens, the premium should rise. It does not on
curated (\cellA{}) or raw (\cellB{}) Python queries (a curated rise
above $0.05$ is excluded). It does rise on TypeScript, and on Rust
without a resolved interval. Instead both retrievers collapse, and the
encoder loses a third to almost half of its \ndcg{10}.
Most replaced tokens are ordinary words, so the edit may also remove
task meaning (\cref{app:genarm-moved}).

The premium thus sits on queries whose rarest token shared with the
code is still common. $\tau$ shows where the premium is, but it is not a
lever: removing the rare tokens lowers both retrievers and, on Python
queries, leaves the premium unchanged.

\subsection{Query Source and Vintage}
\label{sec:r8}
\label{sec:f18}
\paragraph{Pools differ little once labeled alike.} At $c=1{,}000$ every
pool shows a premium, $+0.121$ curated against $+0.128$ to $+0.191$ raw
(\tab{r8}); the \cellB{}-minus-curated gap of $+0.031$ has a
repository-block interval covering zero. Keeping the raw pools' edited
tests as gold more than doubles that gap, so most of the pooled
contrast came from the label rule.

\paragraph{Raw wording raises the premium within a task.} A
$36$-repository replication pairs each raw PR with an LLM rewrite of
the same change, against one gold set and corpus. The raw view's
premium is larger by $+0.091$ for \texttt{jina-code-0.5b} and $+0.111$
for \texttt{Qwen3-Embedding-0.6B}, both intervals excluding zero
(\cref{sec:scaleup}). A date effect of similar size cannot be excluded
there. DeepSWE task descriptions, which describe work to do
rather than a finished change, show no premium distinguishable from
zero (\cref{sec:external}).

\paragraph{Vintage shows no consistent effect.} The 2022, 2023 and 2026
raw pools lie within $0.063$ of each other, with no ordering by date.
On $11$ further Python repositories, 2022 and 2026 PRs differ by
$+0.019$, not the smaller post-cutoff premium that contamination would
predict. With so few repositories, modest effects of wording or date
cannot be excluded.

Wording therefore moves the premium when the task is held fixed; pool origin and date show no consistent effect once labels
match.

\begin{table}[t]
\centering
\footnotesize
\setlength{\tabcolsep}{3pt}
\begin{sc}
\begin{adjustbox}{max width=\columnwidth}
\begin{tabular}{lllrrr}
\toprule
Pool & Query & Date & $n$ & $\Delta$ source & $\Delta$ raw tests kept \\
\midrule
\cellA{}  & curated & pre       & 472 & $+0.121$ & $+0.121$ \\
\cellRA{} & raw     & 2023      & 138 & $+0.128$ & $+0.192$ \\
\cellRB{} & raw     & 2022      & 367 & $+0.191$ & $+0.245$ \\
\cellB{}  & raw     & 2026      & 288 & $+0.152$ & $+0.192$ \\
\bottomrule
\end{tabular}
\end{adjustbox}
\end{sc}
\caption{Premium at $c=1{,}000$ by pool under source-file labels and with the
raw pools' edited test files kept as gold (indexed files only). Curated
labels are unchanged in this comparison, since its gold patch has no
tests. Keeping tests raises every raw pool by $0.04$--$0.06$.}
\label{tab:r8}
\end{table}

\subsection{Predicting the Winning Retriever}
\label{sec:predictable}

\paragraph{The headroom is real.} BM25 strictly beats dense retrieval on
$24\%$ of \cellA{} and $21\%$ of \cellB{} queries, and a per-query
oracle would add about $0.09$ \ndcg{10} over dense alone.

\paragraph{No classifier or LLM judge we tried finds those queries.} We
tried classifiers on structural query features, logistic regression on
query embeddings and two zero-shot LLM judges (\texttt{qwen35-122B},
\texttt{gemma-4-31B-it}). All reach AUROC $0.47$--$0.54$, with every
interval covering $0.5$ (\cref{tab:router-pred}). Once tuned, each falls back to
(nearly) always choosing dense; untuned, the judges pick keyword search
for most queries and lose over $0.1$ \ndcg{10}. The \cellB{} intervals are wide
($n=132$), so a modest signal on raw queries cannot be excluded
(\cref{sec:lexical-rescue}).

BM25 thus wins about a quarter of queries, but nothing we tried can
tell which ones from the query.

\subsection{The Agent Boundary}
\label{sec:agentsummary}
\paragraph{Swapping the search tool changes little; swapping the LLM
changes a lot.} With the LLM fixed, a SWE-agent on the held-out pool
($96$ tasks) scores within about $0.04$ exact-match whether it gets an
unrestricted shell with \texttt{grep}, a lexical index, a dense index
or a $\widehat\tau$-router. Exact-match means the files its final diff edits
equal the gold set. Across four setups of three LLMs on a $30$-task
subset, exact-match ranges from $0.03$ to $0.30$. No tests are run, so this measures localization,
not fix correctness. All runs use clean environments with the merged
fix unreachable (\cref{sec:agent,sec:agent-contamination}).

For agents, then, the choice of LLM matters far more for localization
than the search tool it calls.

\section{Discussion and Conclusion}
\label{sec:discussion}
\label{sec:conclusion}

\paragraph{Reconciling the debate.} Benchmark gaps and agent
experience can both be right. Against BM25 with the dense retriever's
chunking, or with edited tests counted as relevant, dense encoders lead
by $0.1$--$0.2$ \ndcg{10}. Against whole-file BM25 on source-file
labels they lead by little, mostly in coverage. Where they still lead,
the query names no rare identifier. A language model can add those
identifiers, and \texttt{grep}-style search then matches the encoder.
An agent that picks its own search terms does the same. Fusing the two
retrievers still beats either alone (\cref{tab:budget}).

\paragraph{Guidelines.} \gone Measure dense gains against a fair
lexical baseline: whole files for BM25, templated PR text stripped, and
one stated rule for which edited files count as relevant, applied to
every query source. Report Recall@10 and an anchor-augmented BM25 arm
next to the dense number. \gtwo Keep dense retrieval and fuse it with
whole-file BM25; per-query routing did not beat the fusion. If the
caller is an agent, spend on the model before the search endpoint.
\gthree When a query lacks the code's vocabulary, add LLM-written
anchors before searching. The call recurs per query, so a dense index
is cheaper for repositories searched often (\cref{app:genarm-moved}).

We release \benchmark{} with source and test labels, the
PR-boilerplate stripper, and the generation and agent-evaluation
harnesses.

\section*{Limitations}

The core covers ten Python repositories and the cross-language PR pools
one repository per language, so modest effects cannot be excluded. BM25 was tuned on the
evaluation queries, generation results depend on the generator, and
labels are the files one change edited. Function- and line-level
targets are untested.

\bibliography{custom}

\clearpage
\appendix

\section{Extended Related Work}
\label{app:related-ext}

\paragraph{Benchmarks.} Agent suites refresh or harden SWE-bench
\citep{zhang2025swebenchlive, deng2025swebenchpro, huang2026deepswe,
merrill2026terminalbench} but score the whole agent loop, so they cannot
attribute a failure to retrieval. Code-retrieval benchmarks
\citep{husain2019codesearchnet, huang2021cosqa, lu2021codexglue,
li2024coir, xue2026coreb} use docstring, web or competition queries rather
than repository changes. SWE-Explore \citep{zhang2026sweexplore} labels
the line regions that successful agents read. Its explanation of why
lexical search fails on issue text matches what $\tau$ measures: the
issue names no rare identifier of the edited module. Our
encoders descend from code pretraining
\citep{feng2020codebert, guo2021graphcodebert, guo2022unixcoder,
wang2023codet5plus}; the canonical one is \texttt{jina-code}
\citep{guenther2024jinacode}.

\paragraph{Retrieval methods.} Fusion, dense, late-interaction,
learned-sparse and cascaded pipelines combine lexical and dense signals
\citep{cormack2009rrf, karpukhin2020dpr, khattab2020colbert,
santhanam2022colbertv2, formal2021splade, lin2021pyserini,
lin2021densesparse, ma2023bm25dense}; we ask when the dense signal adds
value and whether that can be predicted per query. $\tau$ builds on
pre-retrieval query-performance prediction
\citep{cronen2002predicting, he2006icf, hauff2008preretrieval,
carmel2010estimating, zhou2007qpp, arabzadeh2021bert_qpp}. LLM query
expansion helps BM25 and dense retrieval
\citep{jagerman2023qe, mackie2023grf, wang2023query2doc, gao2023hyde};
we use it as a controlled intervention against pseudo-relevance
feedback \citep{lavrenko2001rm}. LLM systems outside code also rewrite
a user request into search queries before retrieving, for example over
a knowledge base and a web search engine \citep{xue2023weaverbird}.
Rerankers trained on general web data transfer imperfectly to new
domains \citep{xue2026shopranker}; our reranker baseline
(\texttt{jina-reranker-v3}) is such a general-purpose model.

\paragraph{Embeddings beyond code.} A survey of omnimodal embeddings
\citep{zhang2026omnimodal} organizes cross-modal unification around
coverage, comparability and compositionality. Domain-specific retrieval
also requires careful evaluation: ZooClaw-FashionSigLIP2
\citep{xue2026fashionsiglip2} studies adaptation to fashion retrieval
and structural biases in benchmark ground truth. These broader settings
motivate examining how benchmark construction shapes measured retrieval
gains; our experiments isolate this question for repository code search.

\paragraph{Practice and contamination.} The claim we test is that
embeddings were not the bottleneck for a \texttt{grep}-driven SWE-bench
agent \citep{augment2025}; code-graph systems argue for structural
retrieval instead \citep{abdelaziz2021codegraph}. Contamination work
\citep{magar2022contamination, sainz2023nlpevaltrouble} motivates our
post-cutoff pools and the exposure test of generation.

\section{Benchmark Construction Details}
\label{app:benchmark-details}

The construction keeps the pools comparable. Every pool uses one corpus
snapshot per repository and one label rule (the indexed, non-test files
a change edited), and every raw pool uses the same text filters. The
checks below show that the snapshot, unlabeled relevant files and
BM25's tokenizer do not drive the results, but the test-file rule does.

\subsection{Construction Details by Step}
\label{sec:construction}

These notes add to Steps 1--6 of \cref{sec:pipeline} only what the
main text omits.

\paragraph{Repositories.} SWE-bench Verified's $500$ instances cover
twelve repositories; \texttt{seaborn} and \texttt{flask} are too small
for raw pools, leaving ten with enough merged PRs in every window. The
core is Python because SWE-bench Verified is; Multi-SWE-bench enters as
an \extset{} pool (\cref{app:curated-multiling}), since its Python split
repackages SWE-bench Verified.

\paragraph{Mining and labels.} Raw PRs come from the GitHub search API,
most recent first, capped at $100$ per repository ($40$ for \cellRA{});
the query is the PR title and body. Gold files come from the SWE-bench
gold patch (\cellA{}) or the PR's changed-file list (raw pools). The
gold patch has no tests. PR file lists, by contrast, include tests,
documentation, changelogs and configuration files, so scored gold sets
keep only indexed, non-test files (\cref{sec:test-gold}). The release
flags each edited file as source, test or non-indexed. In the raw
pools, tests are $29$--$33\%$ of edited files and appear in
$58$--$63\%$ of queries. Unindexed documentation, changelogs and
templates are $29$--$32\%$ and appear in about half of queries. \cellRB{} was added because
\cellRA{}, mined at $40$ PRs per repository, is small.

\paragraph{Filtering.} Filters apply in the order of \cref{tab:funnel}.
The gold rule asks for one to three scored gold files, all present at
the snapshot, so changes that add their source file are excluded.
Edited tests and documentation do not count toward the cap and need not
exist. The cap keeps the task a localization problem. The text rule
drops bodies under $100$ characters, PRs touching more than $20$ files
and docs-, CI- or test-only changes. Paths
under \texttt{src/X} map to \texttt{X} when only the latter exists.
Analyses that needed new retrieval runs use the $461$ raw queries of
the original, stricter rule. The queries added by the unified rule have
a lower premium at $c=1{,}000$, but no detectable difference in the
tuned premium or the generation comparison. \texttt{sphinx} yields no \cellB{}
query: it merged ten PRs in the 2026 window.

\begin{table*}[t]
\centering
\small
\setlength{\tabcolsep}{4pt}
\begin{sc}
\begin{adjustbox}{max width=\textwidth}
\begin{tabular}{lrrrrrrrrrrr}
\toprule
 & & \multicolumn{6}{c}{Dropped by text rule} & \multicolumn{3}{c}{Dropped by gold rule} & \\
\cmidrule(lr){3-8}\cmidrule(lr){9-11}
Pool & Candidates & $<$100 ch. & $>$20 files & docs & CI & tests & no code & no source & $>$3 files & missing & Kept \\
\midrule
\cellA{}  & 497 & -- & -- & -- & -- & -- & -- & 0 & 10 & 15 & 472 \\
\cellB{}  & 897 & 192 & 19 & 82 & 107 & 49 & 49 & 24 & 32 & 55 & 288 \\
\cellRA{} & 400 & 109 & 4 & 41 & 49 & 17 & 22 & 10 & 8 & 2 & 138 \\
\cellRB{} & 939 & 251 & 27 & 90 & 63 & 25 & 48 & 12 & 41 & 15 & 367 \\
\midrule
Total & 2{,}733 & 552 & 50 & 213 & 219 & 91 & 119 & 46 & 91 & 87 & 1{,}265 \\
\bottomrule
\end{tabular}
\end{adjustbox}
\end{sc}
\caption{Quality-filter funnel for the four core pools, ten repositories.
Text rules apply to raw PRs only. The gold rule counts scored gold files
(indexed, non-test): \emph{no source} means that no scored gold file
remains, and \emph{missing} means that at least one is absent at the
pinned snapshot (\cellRA{}: the 2024-Q1 snapshot).}
\label{tab:funnel}
\end{table*}

\paragraph{Preprocessing.} Raw PR descriptions carry ticket headers,
checklists, AI-disclosure sections, checkbox lines and HTML comments;
unstripped, BM25 on \texttt{django} \cellB{} collapses to
\ndcg{10}$=0.010$. The stripper (\cref{app:stripper}) shortens raw
queries by about $40\%$ and curated statements by about $4\%$.

\paragraph{Corpus.} Each repository is pinned at the base commit of its
most recent SWE-bench Verified instance, so every \cellA{} base commit
precedes the snapshot. The corpus is every \texttt{.py} file, tests
included (\cref{tab:repostats}). \cellRA{} uses a 2024-Q1 snapshot. As
a result, most curated fixes are already in the snapshot, and the
\cellB{} changes come about three years after the indexed code.
``Post-cutoff'' therefore describes the query and the change but not
the code. \Cref{app:percommit} tests whether this matters.

\begin{table*}[t]
\centering
\small
\begin{sc}
\begin{adjustbox}{max width=\textwidth}
\begin{tabular}{lrrrrrr}
\toprule
\textbf{Repo} & \textbf{Snapshot} & \textbf{Corpus files} & \textbf{\cellA{}} & \textbf{\cellB{}} & \textbf{\cellRA{}} & \textbf{\cellRB{}} \\
\midrule
\texttt{django}        & 2023-07 & 2{,}752 & 226 & 45 & 10 & 24 \\
\texttt{sympy}         & 2023-02 & 1{,}511 &  72 & 57 & 33 & 69 \\
\texttt{scikit-learn}  & 2023-05 &   920   &  18 & 36 & 15 & 52 \\
\texttt{astropy}       & 2023-06 &   899   &  21 & 14 & 15 & 28 \\
\texttt{matplotlib}    & 2023-08 &   895   &  34 & 13 & 15 & 44 \\
\texttt{sphinx}        & 2023-07 &   581   &  44 &  0 & 17 & 31 \\
\texttt{pytest}        & 2022-10 &   244   &  19 & 34 &  7 & 23 \\
\texttt{xarray}        & 2022-12 &   174   &  22 & 30 & 15 & 46 \\
\texttt{pylint}        & 2023-07 & 2{,}189 &   8 & 32 &  8 & 35 \\
\texttt{requests}      & 2022-01 &    35   &   8 & 27 &  3 & 15 \\
\midrule
\textbf{Total}         &         & 10{,}200 & 472 & 288 & 138 & 367 \\
\midrule
Mean edited files (as mined) & & & 1.15 & 2.96 & 3.22 & 2.69 \\
Mean gold files (scored) & & & 1.15 & 1.17 & 1.36 & 1.21 \\
Queries editing a test & & & 0\% & 63\% & 59\% & 58\% \\
Mean query length (ch.) & & & 1{,}606 & 662 & 472 & 493 \\
\bottomrule
\end{tabular}
\end{adjustbox}
\end{sc}
\caption{Per-repository corpus and query counts after quality filtering.
\emph{Snapshot}: creation month of the SWE-bench Verified instance whose
base commit pins the corpus (\cellRA{} uses a 2024-Q1 snapshot). Query
length is in characters after stripping.}
\label{tab:repostats}
\end{table*}

\subsection{Per-Task Pre-Change Snapshots}
\label{app:percommit}

The shared snapshot does not drive the premium. We re-score $933$ core
queries at each task's own base commit ($834$ commits). Retrieval is
otherwise identical, and a guard reproduces the shared-snapshot results
exactly. No tuned premium moves by more than $0.025$, and the
raw-minus-curated gap is unchanged (\cref{tab:percommit}). The curated
premium rises slightly, so a snapshot that already contains the fix
favors BM25 a little on curated queries.

\begin{table}[t]
\centering
\footnotesize
\setlength{\tabcolsep}{3pt}
\begin{sc}
\begin{adjustbox}{max width=\columnwidth}
\begin{tabular}{lrrrr}
\toprule
Pool & $n$ & Shared snapshot & Own base commit & $\Delta$ \\
\midrule
\cellA{}  & 472 & $+0.034$ {\scriptsize[$-0.024$, $+0.117$]} & $+0.056$ {\scriptsize[$+0.024$, $+0.116$]} & $+0.022$ \\
\cellB{}  & 132 & $+0.074$ {\scriptsize[$-0.027$, $+0.220$]} & $+0.098$ {\scriptsize[$-0.003$, $+0.239$]} & $+0.024$ \\
\cellRA{} &  72 & $+0.124$ {\scriptsize[$-0.053$, $+0.309$]} & $+0.102$ {\scriptsize[$-0.062$, $+0.347$]} & $-0.022$ \\
\cellRB{} & 255 & $+0.112$ {\scriptsize[$+0.028$, $+0.240$]} & $+0.125$ {\scriptsize[$+0.033$, $+0.259$]} & $+0.013$ \\
\bottomrule
\end{tabular}
\end{adjustbox}
\end{sc}
\caption{Tuned premium (dense minus whole-file BM25, \ndcg{10}) with each
query scored at the shared snapshot and at its own pre-change commit,
same queries, repository-clustered 95\% intervals. \cellRA{}'s shared
corpus is its 2024-Q1 snapshot; two \cellRB{} tasks whose gold is absent
at the base commit are dropped.}
\label{tab:percommit}
\end{table}

\subsection{Label Definition and the Test-File Bias}
\label{sec:test-gold}

Counting edited test files as relevant rewards a habit of dense
retrievers, not localization. If one rule is used for all pools,
either source files only or all edited files, the curated and raw
premiums are indistinguishable. Only mixing the two separates them
(\cref{tab:test-gold}).

\paragraph{Rule.} A file is a test if its path contains a test directory
(\texttt{test}, \texttt{tests}, \texttt{testing}, \texttt{\_\_tests\_\_},
\dots) or its name matches a test pattern (\texttt{test\_*.py},
\texttt{conftest.py}, \texttt{*\_test.go}, \texttt{*.spec.ts}, \dots); it
is indexed if the pool's corpus indexes its extension. Every analysis is
re-scored from stored top-10 lists with these files removed from the
gold set; retrieved tests then count as non-relevant.

\begin{table}[t]
\centering
\footnotesize
\setlength{\tabcolsep}{3pt}
\begin{sc}
\begin{adjustbox}{max width=\columnwidth}
\begin{tabular}{llrrl}
\toprule
BM25 & Labels & \cellA{} & \cellB{} & Gap [95\% CI] \\
\midrule
$c=1{,}000$ & source files   & $+0.121$ & $+0.175$ & $+0.054$ $[-0.045,+0.198]$ \\
         & all edited     & $+0.191$ & $+0.204$ & $+0.013$ $[-0.051,+0.122]$ \\
         & mixed (as mined) & $+0.121$ & $+0.204$ & $+0.083$ \\
\midrule
tuned    & source files   & $+0.034$ & $+0.074$ & $+0.040$ $[-0.059,+0.176]$ \\
         & all edited     & $+0.165$ & $+0.152$ & $-0.013$ $[-0.122,+0.096]$ \\
         & mixed (as mined) & $+0.034$ & $+0.152$ & $+0.118$ \\
\bottomrule
\end{tabular}
\end{adjustbox}
\end{sc}
\caption{Premium (canonical encoder minus BM25, \ndcg{10}) under three
label rules, on the original \cellB{} queries (\cref{sec:methods});
gaps are \cellB{} minus \cellA{}, repository-block intervals. On the
full raw pools the \cellB{} tuned premium is $+0.059$ (source files)
and $+0.136$ (all edited files).}
\label{tab:test-gold}
\end{table}

\paragraph{Dense retrievers rank tests first far more often.} Dense
retrieval puts a test file first for $59\%$ of curated and $28$--$35\%$
of raw queries, against $34\%$ and $12$--$16\%$ for tuned BM25. Word
overlap and file names do not explain this: those tests share fewer
query words than the source files dense retrieval ranks, and few are
named after a gold module. Reproduction code in the query strengthens
the preference, but it remains when the query has no code.

\paragraph{Beyond code.} Many relevance labels are harvested from
records made for other purposes, such as a change's file list, a
ticket's links or a clinical note's codes. The harvesting rule then
decides what counts as relevant, and retriever families may differ in
how easily they find the extra items. A retrieval comparison is only as
consistent as its label rule.

\subsection{Audit of Unlabeled Retrieved Files}
\label{app:relevance-audit}

Missing labels lower both retrievers' scores about equally, so they do
not explain the premium. An LLM judge (\texttt{qwen35-122B}) read each
change description together with the top-3 non-gold files from the
canonical encoder and from whole-file BM25. One gold file per query was
mixed in without its label. We used all $132$ original \cellB{} queries and $132$
sampled \cellA{} queries. The judge marked $75$--$78\%$ of gold files as
files to edit, but only $5$--$6\%$ of the non-test misses, with no
difference between the retrievers. The dense retriever's apparent
surplus of editable misses comes from its preference for tests
(\cref{sec:test-gold}).

\subsection{Which Part of the Expansion Matters}
\label{app:anchor-split}

Generation works through the identifiers it writes. We append one group
of generated fields at a time (\cref{tab:anchor-split}). Identifiers,
APIs and error terms close the premium with or without file paths, and
the code skeleton nearly does. Prose search phrases help only through
the identifiers they contain; with those removed, the premium stays
where it was.

\begin{table*}[t]
\centering
\small
\begin{sc}
\begin{adjustbox}{max width=\textwidth}
\begin{tabular}{lrrrrrrr}
\toprule
& & \multicolumn{6}{c}{\textbf{dense $-$ BM25 after adding to the query}} \\
\cmidrule(lr){3-8}
\textbf{Pool} & \textbf{query only} & \textbf{all fields} & \textbf{ident.+APIs+errors} & \textbf{ident.\ + paths} & \textbf{code skeleton} & \textbf{prose phrases} & \textbf{prose, no identifiers} \\
\midrule
\cellA{} & $+0.034$ & $-0.081$ & $-0.040$ & $-0.039$ & $-0.047$ & $-0.020$ & $+0.027$ \\
\cellB{} & $+0.074$ & $-0.052$ & $-0.006$ & $-0.009$ & $+0.018$ & $+0.036$ & $+0.086$ \\
\cellRA{} & $+0.124$ & $-0.030$ & $+0.007$ & $+0.006$ & $-0.008$ & $+0.019$ & $+0.128$ \\
\cellRB{} & $+0.109$ & $-0.099$ & $-0.045$ & $-0.028$ & $-0.001$ & $+0.016$ & $+0.109$ \\
\bottomrule
\end{tabular}
\end{adjustbox}
\end{sc}
\caption{Dense premium (canonical encoder minus whole-file BM25,
\ndcg{10}, source-file labels) after appending one group of generated
fields to the query. Original queries ($n=472$, $132$, $72$, $257$), so
values differ from the full-pool \tab{genarm-tuned} and from
\tab{percommit} ($255$ \cellRB{} tasks). Repository-clustered intervals
are in the released analysis file.}
\label{tab:anchor-split}
\end{table*}

\subsection{Identifier-Splitting BM25}
\label{app:idsplit}

Splitting snake- and camel-case identifiers does not change the
conclusion. It helps BM25 at $c=1{,}000$, but in Python whole files help
more. Against BM25's best configuration the premium is at most about
$+0.09$ in every pool, Python or not (TypeScript already used
splitting).

\subsection{Held-Out Pool}
\label{sec:heldout}

The held-out pool tests the findings on repositories we did not use
during construction: nine Python repositories outside the core and
outside SWE-bench Verified. We mined PRs merged on or after 2025-04-01
and applied the \cellB{} filter ($756$ mined PRs, $238$ usable, $139$
with a source gold file). On the $139$ scored queries the premium is
$+0.068$. It is positive on six of nine repositories and falls across
$\tau$ quartiles as in the core pools. The pool also hosts the agent evaluation (\secref{agent}).

\section{Methods Details}
\label{app:methods-details}

BM25 is BM25Okapi (\texttt{rank\_bm25}) over chunks or whole files,
with chunk scores max-pooled to the file. Dense retrieval uses
\texttt{sentence-transformers} encoders from six families and thirteen
checkpoints: \texttt{jina-code-embeddings-\{0.5b, 1.5b\}},
\texttt{Qwen3-Embedding-\{0.6B, 4B, 8B\}}, \texttt{C2LLM-\{0.5B, 7B\}},
\texttt{F2LLM-\{0.6B, 4B\}}, \texttt{e5-\{base, large\}-v2} and
\texttt{gte-Qwen2-\{1.5B, 7B\}-instruct}. Each chunk is encoded once,
similarity is cosine and chunk scores are max-pooled to files. All
encoders run without instruction prefixes. This uniform protocol
penalizes models that depend on instructions, \texttt{gte-Qwen2} most. Per
query we report \ndcg{10} and Recall@10 at file granularity, the
premium $\Delta(q)=\ndcg{10}_\text{dense}(q)-\ndcg{10}_\text{BM25}(q)$
and $\tau(q)$ (\secref{tau}).

\paragraph{Statistics.} Bootstraps use $10{,}000$ resamples and fixed
seeds. Per-query comparisons use a paired bootstrap over per-query
\ndcg{10}. Cross-slice analyses resample repositories, then tasks, and
keep all encoders' results for a task together. For gaps between pools,
repository-block intervals are primary; query-level intervals are
conditional on the $19$ repositories pooled. An interval that includes
zero is not evidence of no effect; outside the preset $\pm0.05$ margin
we report attenuation, not equivalence.

\section{Extended Findings}
\label{app:findings-ext}

Three results support the main text. BM25 is usable on raw text only
after PR templates are stripped. The $\tau$-effect is weak per query but
large across query groups, and it is exploratory. Under one label
definition, the query pools are hard to tell apart.

\subsection{PR-Boilerplate Stripping Recovers BM25}
\label{sec:f5}

Templates break BM25 but not dense retrieval, so queries taken from PRs
or issues must be stripped. On raw \cellB{}
queries stripping roughly halves the premium ($+0.369$ to $+0.175$), entirely through
BM25 gains. Without stripping, BM25 on \texttt{django} returns the same template-matching files (\texttt{global\_settings.py},
\texttt{docs/conf.py}) for almost every query. The repeated headers,
checklists and HTML comments dominate BM25's IDF-weighted query, while
the encoder largely ignores them. A hand audit of eight stripped queries
confirms that titles and substantive prose survive.

\subsection{The $\tau$-Effect in Detail}
\label{app:tau-ext}

$\tau$ predicts how large the premium is for a group of queries, not
for a single query, and the evidence for it is exploratory.

\paragraph{$\tau$ is MaxIDF on the tokens a query shares with its
target.} $\tau(q)$ (\cref{eq:tau}) is the MaxIDF predictor of
\citet{hauff2008preretrieval} restricted to tokens shared with the
first indexed, non-test gold file. We chose it among four variants
because it transfers best across slices (\tab{r3}); that choice was made
on the same slices. Its query-only analogue $\widehat\tau$ carries
little per-query signal.

\paragraph{The premium falls with $\tau$ in both cells.} Pooled quartile
means fall monotonically, from $+0.217$ to $+0.050$; the raw cell decays
more steeply, and the curated cell has one interior wobble.

\paragraph{The association is weak per query but large per group.}
The $\tau$ quartile explains about $2\%$ of per-query variance, as much
as knowing the repository; most of that variance is metric noise. Yet
the drop from the lowest to the highest quartile is as large as the
whole spread of the premium across encoders. $\tau$ thus shows where
the premium is but cannot route single queries
(\secref{lexical-rescue}).

\subsection{Query Source and Vintage in Detail}
\label{app:q2-ext}

Under one label rule the core pools are not distinguishable,
vintage has no consistent effect, and the raw-PR premium recurs in
other languages.

\paragraph{The pooled curated--raw gap was mostly a label contrast.}
With source-file labels the raw-minus-curated gap is $+0.031$ and not
significant under repository-block resampling; keeping the raw pools'
edited tests more than doubles it (\cref{sec:test-gold}). The pools
also differ in repositories, query length and gold-set size, so even
the residual gap is descriptive. Wording does matter when the task is
held fixed (\cref{sec:scaleup}).

\paragraph{Vintage has no consistent effect.} On $11$ further Python
repositories, we score PRs from 2022 and from 2026 against one pinned
snapshot per repository. The date contrast has both signs across
repositories and pools; pooled, it is $+0.019$. Contamination would predict a
smaller post-cutoff premium, which we do not see, though a modest
effect is not excluded.

\paragraph{The raw-PR premium recurs outside Python.} At $c=1{,}000$ raw
PRs keep a premium in JavaScript, TypeScript, Go and Rust; against the
strongest BM25 we ran it remains on TypeScript and Rust but not on Go
(\cref{app:idsplit}). Each language is a single repository.

\section{Generation and Anchor Ablation}
\label{app:genarm-ext}
\label{app:genarm-moved}

This appendix gives details for the two query interventions of
\cref{sec:genarm} and \cref{sec:tau}; the tables below hold the
numbers.

\paragraph{Setup.} One generator (\texttt{qwen35-122B}, temperature $0$,
one frozen prompt) reads only the stripped query, never gold paths or
patch content. It writes identifiers, file and module hints, APIs,
error terms and a code skeleton, which we prepend to the query for
BM25. A control arm reproduces the cached BM25 exactly.

\paragraph{Generation does more than tuning BM25.} The anchors help BM25
more than any chunk size does. Over whole files, generate-then-grep
matches the canonical encoder on every raw pool and beats it on curated
queries (\tab{genarm-tuned}). The residuals cover zero but are not
shown to be equivalent to zero. We therefore say that the premium is
removed, but not that the two retrievers are equal.

\paragraph{The anchors carry evidence, not just more text.}
Corpus-based expansion (RM3) lowers BM25 rather than raising it.
Embedding the expanded query helps the encoder far less than it helps
BM25, and SPLADE and ColBERTv2 gain less from the same anchors
(\tab{gen-controls}, \tab{encoder2}). What helps is specific
identifiers that a lexical index can match exactly.

\paragraph{A stronger encoder does not change the picture.}
\texttt{Qwen3-Embedding-8B} keeps a larger premium over the tuned BM25
on the raw pools, but generate-then-grep again matches or beats it in
every pool (\tab{encoder2}).

\paragraph{The size of the effect depends on the generator.}
\label{app:genarm}
Across five generators from $4$B to $122$B the direction always holds,
but only the largest turns the curated premium negative.
\texttt{gemma-4-31B-it} and \texttt{deepseek-v4-flash} close about
$80\%$ of it; \texttt{Qwen3-8B} closes a little over half against the
canonical encoder and about two thirds against the encoder built from
its own weights.

\paragraph{Memorization does not explain it.} Generation also removes
the premium on post-cutoff PRs that appear in no benchmark, and the
effect survives when every file reference is stripped from the anchors.
On held-out PRs opened after the generator's release ($116$ queries,
seven repositories), generate-then-grep still beats dense retrieval,
with an interval that excludes zero. Naming the gold file is not needed
either. The generator names it more often for curated than for raw
queries, yet \cellRB{} closes its whole premium although the gold file
is named for only half its queries.

\paragraph{Generation is cheaper for rarely searched repositories.} One
\texttt{Qwen3-8B} call takes about $1.5$\,s per query. The dense index
needs a one-off encoding pass per repository (about four minutes for
\texttt{django}). Generation is therefore cheaper below roughly $158$
queries per repository. Failed generations fall back to the original
query, which biases against generation.

\paragraph{Removing anchors hurts both retrievers.} We replace the five
rarest tokens a query shares with its gold file by synthetic strings of
the same length and case pattern. The placebo replaces the same number
of equally rare tokens absent from the gold file, and a guard arm
reproduces the cached baseline exactly. If dense retrieval matched
meaning without those tokens, its premium should rise. On curated and
raw Python queries it does not: both retrievers lose a third to almost
half of their \ndcg{10}, by similar amounts. The premium does rise on
TypeScript, and on Rust without a resolved interval. Most replaced
tokens are ordinary words rather than identifiers. Our attempt at
meaning-preserving substitutes failed its fidelity check. The edit thus
rules out matching that ignores these tokens, but it cannot tell the
loss of anchors apart from a loss of task meaning. The evidence for the mechanism therefore
comes from adding anchors. An earlier version based on LLM
rewrites (\cellC{}) removes too much of the query to be informative.

\begin{table}[t]
\centering
\small
\setlength{\tabcolsep}{2.5pt}
\begin{sc}
\begin{adjustbox}{max width=\columnwidth}
\begin{tabular}{lrrrrrrr}
\toprule
& \multicolumn{3}{c}{\textbf{Qwen3-Emb-8B}} & \multicolumn{2}{c}{\textbf{SPLADE}} & \multicolumn{2}{c}{\textbf{ColBERTv2}} \\
\cmidrule(lr){2-4}\cmidrule(lr){5-6}\cmidrule(lr){7-8}
\textbf{Cell} & \textbf{prem.} & \textbf{after gen.} & \textbf{gap} & \textbf{query} & \textbf{+anch.} & \textbf{query} & \textbf{+anch.} \\
\midrule
\cellA{} & $+0.017$ & $-0.099$ & $+0.070$ & 0.390 & 0.436 & 0.397 & 0.428 \\
\cellB{} & $+0.054$ & $-0.072$ & $+0.064$ & 0.540 & 0.557 & 0.422 & 0.570 \\
\cellRA{} & $+0.120$ & $-0.034$ & $+0.076$ & 0.523 & 0.582 & 0.454 & 0.542 \\
\cellRB{} & $+0.159$ & $-0.049$ & $+0.162$ & 0.489 & 0.547 & 0.421 & 0.541 \\
\bottomrule
\end{tabular}
\end{adjustbox}
\end{sc}
\caption{Generality checks (\ndcg{10}). Left: the generative-arm analysis
repeated with \texttt{Qwen3-Embedding-8B} (premium against whole-file
BM25; residual against whole-file BM25 on the query plus anchors; gap =
BM25's gain from the anchors minus the encoder's). Its original-query
lists reproduce the cached canonical runs exactly on \cellA{} and
\cellB{}. On the full raw pools, Qwen3-Emb-8B's tuned premium is
$+0.037$, $+0.117$ and $+0.146$ (\cellB{}, \cellRA{}, \cellRB{}) and its
residual after generation $-0.075$, $-0.031$ and $-0.041$. Right: off-the-shelf SPLADE (\texttt{splade-cocondenser-ensembledistil})
and ColBERTv2 (exhaustive MaxSim, queries up to 256 tokens) on $c=1{,}000$
chunks max-pooled per file, on the original and expanded query; raw
pools: original queries ($132$ \cellB{}, $72$ \cellRA{}, $257$ \cellRB{};
\cref{sec:methods}), so
values differ from the full-pool \tab{genarm-tuned}. For
reference, on these queries whole-file BM25 scores $0.498$, $0.539$, $0.399$, $0.463$ and
the canonical encoder $0.532$, $0.613$, $0.523$, $0.572$. Source-file
labels.}
\label{tab:encoder2}
\end{table}

\begin{table}[t]
\centering
\small
\setlength{\tabcolsep}{3pt}
\begin{sc}
\begin{adjustbox}{max width=\columnwidth}
\begin{tabular}{lrrrrrr}
\toprule
& \multicolumn{2}{c}{\textbf{nDCG@10}} & \multicolumn{2}{c}{\textbf{Recall@10}} & \multicolumn{2}{c}{\textbf{MRR@10}} \\
\cmidrule(lr){2-3}\cmidrule(lr){4-5}\cmidrule(lr){6-7}
\textbf{Cell} & \textbf{vs BM25} & \textbf{vs gen.} & \textbf{vs BM25} & \textbf{vs gen.} & \textbf{vs BM25} & \textbf{vs gen.} \\
\midrule
\cellA{} & $+0.034$ & $-0.081$ & $+0.099$ & $-0.001$ & $+0.017$ & $-0.106$ \\
\cellB{} & $+0.059$ & $-0.054$ & $+0.106$ & $+0.014$ & $+0.036$ & $-0.084$ \\
\cellRA{} & $+0.089$ & $-0.059$ & $+0.161$ & $-0.037$ & $+0.062$ & $-0.064$ \\
\cellRB{} & $+0.125$ & $-0.062$ & $+0.188$ & $-0.031$ & $+0.108$ & $-0.071$ \\
\bottomrule
\end{tabular}
\end{adjustbox}
\end{sc}
\caption{The dense advantage under three metrics: the canonical encoder
($c=1{,}000$) minus BM25 over whole files on the original query (``vs
BM25'') and on the query plus generated anchors (``vs gen.''),
source-file labels, full pools. Against plain BM25 the advantage lies
mostly in Recall@10 (significant in the three raw pools); MRR@10
differs significantly only on \cellRB{}. Against generate-then-grep it
disappears on every metric. CIs are in the released analysis files.}
\label{tab:metrics}
\end{table}

\begin{table}[t]
\centering
\small
\setlength{\tabcolsep}{3pt}
\begin{sc}
\begin{adjustbox}{max width=\columnwidth}
\begin{tabular}{lrrrrrl}
\toprule
& \multicolumn{3}{c}{\textbf{BM25, whole files}} & \multicolumn{2}{c}{\textbf{dense}} & \\
\cmidrule(lr){2-4}\cmidrule(lr){5-6}
\textbf{Cell} & \textbf{query} & \textbf{+anch.} & \textbf{RM3} & \textbf{query} & \textbf{+anch.} & \textbf{gain gap [95\% CI]} \\
\midrule
\cellA{}  & 0.498 & 0.613 & 0.429 & 0.532 & 0.566 & $+0.081$ $[+0.029,+0.128]$ \\
\cellB{}  & 0.539 & 0.664 & 0.497 & 0.613 & 0.639 & $+0.099$ $[+0.047,+0.130]$ \\
\cellRA{} & 0.399 & 0.553 & 0.387 & 0.523 & 0.595 & $+0.082$ $[+0.006,+0.129]$ \\
\cellRB{} & 0.463 & 0.671 & 0.439 & 0.572 & 0.640 & $+0.140$ $[+0.087,+0.221]$ \\
\bottomrule
\end{tabular}
\end{adjustbox}
\end{sc}
\caption{Controls for the generative arm (\ndcg{10}). RM3: pseudo-relevance
feedback on the same whole-file BM25 (10 documents, 10 terms, weight
$0.5$, untuned). Dense ``+anch.'': the canonical encoder embeds the
expanded query (HyDE-style). Gain gap: BM25's gain from the anchors minus
dense retrieval's, with repository-clustered 95\% CIs; source-file
labels. Loader guards:
RM3's first pass reproduces whole-file BM25 exactly; dense on the
original query reproduces the canonical lists for all but one \cellRA{}
query. Raw pools: original queries ($n=132$, $72$, $257$;
\cref{sec:methods}), so values differ from the full-pool
\tab{genarm-tuned}.}
\label{tab:gen-controls}
\end{table}

\section{Decision Analysis: Protocol and Lockbox}
\label{app:decision}

The preregistered router failed on the lockbox, while both fixed
fusions, configured on development data alone, held.

\paragraph{What was frozen.} The actions are dense retrieval
(\texttt{jina-code-0.5b}, $c=1{,}000$), whole-file BM25 (the best pooled
chunk size on development data) and their RRF ($k=60$). Development is
\cellA{} and \cellB{} ($10$ repositories), evaluated leave one
repository out. The router uses gradient-boosted regressors over $48$
query, list-agreement and score features; it serves RRF unless another
action is predicted to beat it. Of about $30$ candidates, it was the
only one whose development gain excluded zero. We froze it with its
hash and a written criterion: on the pooled lockbox, the lower 95\%
bound of routed minus always-RRF must be positive. An earlier router had already failed the
same kind of test on $29$ held-out repositories.

\paragraph{The router fails; the fusions hold.} The lockbox ($539$
DeepSWE and cross-language TypeScript, Go and Rust queries, $91$
repositories) was scored once on the labels
as released. The router does not beat always fusing, and the
lockbox is not reused for routing. Both fusions beat dense retrieval,
and re-scoring with source-file labels changes no conclusion.

\paragraph{The fusion gain is not carried by three repositories.}
$80\%$ of lockbox queries come from the TypeScript, Go and Rust slices.
With the $91$ repositories weighted equally (post hoc), both fusions
keep their gain over dense retrieval, even with those three slices
dropped. The generation tier's extra gain over plain fusion does not
survive; it comes from the three single-repository slices.

\section{Data Release and Personal Information}
\label{app:release}
\label{app:ethics}

\benchmark{} redistributes only public GitHub content (problem
statements, PR descriptions and code at pinned SHAs) under the source
repositories' licenses. Released queries pass the same boilerplate
stripper used for evaluation, and a release-time scrubber removes every
email address. \texttt{@}-mentions are kept because they are lexical
tokens; most are bots, package scopes or decorators. We redistribute
no commit histories or non-public data, and make no attempt to link or
profile individuals. Reported numbers were computed on the unredacted
text; BM25 rankings differ for a few released queries. The use of LLMs in this work is described in \cref{app:llm-use}.

\section{Reproducibility and Use of LLMs}
\label{app:llm-use}

\paragraph{Reproducibility.} All pools are built from public GitHub
data at pinned SHAs (\tab{cells}). The code and data releases (links on the first page) contain
the query and qrel files, the PR-boilerplate stripper, the cached
per-query results, and the scripts that produce every table and
figure. It also contains the agent-evaluation harness: the
clean-environment builder, \texttt{PATH} shims and bypass auditor
(\secref{agent-contamination}). Bootstrap intervals use $10{,}000$ resamples with fixed seeds;
all encoders are open-weight checkpoints pinned by revision.

\paragraph{Use of large language models.} LLMs are part of the
experiments. They drive the agent evaluation (\secref{agent}) and
generate the query anchors of \secref{genarm}, an experimental
condition that never enters the corpus or the qrels. They also write
the \cellC{} rewrites (\cref{app:cellc}) and produce and check the
curated views of the matched replication (\secref{scaleup}). No LLM
generated, labeled or filtered the queries or qrels of the four core
pools. Coding assistants were used to implement analysis tooling, run
experiments and draft revisions; the authors reviewed and verified all
results and text.

\section{Chunk-Size and $\tau$-Definition Ablations}
\label{sec:robustness}

Chunk size decides the baseline. BM25 keeps improving up to whole files
while the encoder does not, so the common $c=1{,}000$ protocol favors
dense retrieval. Against BM25 at its own best size, no encoder keeps a
significant premium. Among $\tau$ variants, MaxIDF is the most
informative; whole-query predictors carry no signal.

\paragraph{BM25 needs large windows; the encoder does not.} BM25 rises
with chunk size: it levels off from $c=4{,}000$ on \cellA{} and peaks at
whole files on \cellB{}. The canonical encoder stays flat because it
already truncates long inputs (\tab{r5}). Only the canonical encoder
was swept.

\paragraph{Rebasing removes every significant premium.} Scored against
BM25 at its best size, no encoder of thirteen keeps a significantly
positive premium on either pool, and seven fall significantly below zero
on \cellA{} (\tab{chunk-rebase}). BM25's size was picked on the
evaluation queries, so these rebased premiums are lower bounds for dense
retrieval.

\paragraph{MaxIDF is the most informative $\tau$.} Of four shared-token
variants, MaxIDF has the largest gap between the lowest and highest
$\tau$ quartile and transfers best across slices (\secref{law}). Query
length and gold-set size carry no signal, so what matters is
restricting $\tau$ to tokens shared with the target file (\tab{r3}).

\begin{table*}[t]
\centering
\small
\begin{sc}
\begin{adjustbox}{max width=\textwidth}
\begin{tabular}{lrrrl}
\toprule
\textbf{$\tau$ / QPP variant} & \textbf{Pearson $r$} & \textbf{Spearman $\rho$} & \textbf{$Q_1$--$Q_4$} & \textbf{Monotone?} \\
\midrule
$\tau_\text{MaxIDF}$ (chosen)  & $-0.124$ & $-0.105$ & $+0.166$ & yes \\
$\tau_\text{mean5}$            & $-0.050$ & $-0.080$ & $+0.099$ & no \\
$\tau_\text{jaccard}$          & $-0.113$ & $-0.120$ & $+0.136$ & yes \\
$\tau_\text{mean-all}$         & $-0.083$ & $-0.068$ & $+0.076$ & yes \\
\midrule
\multicolumn{5}{l}{\emph{Additional baselines (gold-file count requires labels)}} \\
$|q|$ (query length in tokens) & $+0.040$ & $+0.045$ & $-0.026$ & no \\
$n_\text{gold}$ (gold-file count) & $+0.046$ & $+0.079$ & -- & no \\
\bottomrule
\end{tabular}
\end{adjustbox}
\end{sc}
\caption{Alternative $\tau$ definitions and query-side baselines on the
pooled \cellA{} and original \cellB{} queries ($n=604$, source-file
labels). $Q_1$--$Q_4$: the drop in mean premium from the lowest to the
highest quartile of each variant. MaxIDF has the largest gap; whole-query
predictors carry no comparable signal.}
\label{tab:r3}
\end{table*}

\begin{table*}[t]
\centering
\small
\begin{sc}
\begin{adjustbox}{max width=\textwidth}
\begin{tabular}{lrrrrrrrr}
\toprule
\textbf{$c$ (chars)} & 1{,}000 & 1{,}500 & 2{,}000 & 3{,}000 & 4{,}000 & 6{,}000 & 10{,}000 & file \\
\midrule
BM25 \ndcg{10} (\cellA{}) & 0.411 & 0.427 & 0.443 & 0.465 & 0.488 & 0.494 & \textbf{0.500} & 0.498 \\
dense \ndcg{10} (\cellA{}) & 0.532 & 0.522 & 0.531 & 0.530 & -- & -- & -- & -- \\
BM25 \ndcg{10} (\cellB{}) & 0.438 & 0.466 & 0.469 & 0.475 & 0.461 & 0.485 & 0.496 & \textbf{0.539} \\
\bottomrule
\end{tabular}
\end{adjustbox}
\end{sc}
\caption{Chunk-window sweep (\ndcg{10}, source-file labels; \cellA{}
$n=472$, \cellB{} $n=132$). ``File'' scores each file as one window.
Dense is the canonical encoder, swept on \cellA{} only. Bold: BM25's
best size per pool.}
\label{tab:r5}
\end{table*}

\begin{table*}[t]
\centering
\small
\begin{sc}
\begin{adjustbox}{max width=\textwidth}
\begin{tabular}{lrrlrrl}
\toprule
& \multicolumn{3}{c}{\cellA{} (curated)} & \multicolumn{3}{c}{\cellB{} (raw)} \\
\cmidrule(lr){2-4}\cmidrule(lr){5-7}
\textbf{Encoder} & \textbf{@1{,}000} & \textbf{tuned} & \textbf{repo CI} & \textbf{@1{,}000} & \textbf{tuned} & \textbf{repo CI} \\
\midrule
jina-code-1.5b & $+0.124$ & $+0.035$ & $[-0.026,+0.075]$ & $+0.192$ & $+0.091$ & $[-0.009,+0.230]$ \\
jina-code-0.5b (canonical) & $+0.121$ & $+0.032$ & $[-0.037,+0.103]$ & $+0.175$ & $+0.074$ & $[-0.027,+0.219]$ \\
Qwen3-Emb-8B & $+0.103$ & $+0.015$ & $[-0.021,+0.085]$ & $+0.155$ & $+0.054$ & $[-0.023,+0.174]$ \\
e5-base-v2 & $+0.058$ & $-0.030$ & $[-0.105,+0.014]$ & $+0.084$ & $-0.017$ & $[-0.072,+0.079]$ \\
Qwen3-Emb-4B & $+0.051$ & $-0.038$ & $[-0.070,+0.018]$ & $+0.105$ & $+0.004$ & $[-0.075,+0.127]$ \\
gte-Qwen2-7B & $+0.049$ & $-0.040$ & $[-0.135,+0.003]$ & $+0.029$ & $-0.072$ & $[-0.170,+0.056]$ \\
F2LLM-0.6B & $+0.041$ & $-0.048$ & $[-0.095,-0.006]$ & $+0.091$ & $-0.010$ & $[-0.068,+0.090]$ \\
e5-large-v2 & $+0.035$ & $-0.054$ & $[-0.124,-0.017]$ & $+0.109$ & $+0.009$ & $[-0.069,+0.101]$ \\
F2LLM-4B & $+0.016$ & $-0.073$ & $[-0.135,-0.015]$ & $+0.063$ & $-0.038$ & $[-0.114,+0.070]$ \\
Qwen3-Emb-0.6B & $+0.008$ & $-0.081$ & $[-0.112,-0.023]$ & $+0.100$ & $-0.001$ & $[-0.063,+0.089]$ \\
C2LLM-7B & $-0.042$ & $-0.130$ & $[-0.202,-0.072]$ & $+0.012$ & $-0.089$ & $[-0.188,+0.026]$ \\
C2LLM-0.5B & $-0.043$ & $-0.132$ & $[-0.178,-0.106]$ & $-0.025$ & $-0.126$ & $[-0.225,+0.012]$ \\
gte-Qwen2-1.5B & $-0.166$ & $-0.254$ & $[-0.314,-0.227]$ & $-0.078$ & $-0.178$ & $[-0.291,-0.031]$ \\
\bottomrule
\end{tabular}
\end{adjustbox}
\end{sc}
\caption{Premium against BM25 at $c=1{,}000$ and at its own best chunk size
(``tuned'': $c=10{,}000$ on \cellA{}, whole files on \cellB{}); dense at
$c=1{,}000$ for every encoder; repository-block $95\%$ CIs on the tuned
premium; source-file labels; \cellB{}: the $132$ original queries
(\cref{sec:methods}). Against a tuned BM25 no encoder keeps a
significantly positive premium on either pool.}
\label{tab:chunk-rebase}
\end{table*}

\section{The PR-Boilerplate Stripper}
\label{app:stripper}

The stripper processes the raw PR description line by line. It deletes
HTML comments left by the PR template and drops \emph{entire} markdown
sections whose header matches a boilerplate name (``Trac ticket
number'', ``AI Assistance Disclosure'', ``Checklist'', \dots). It
removes the \texttt{\#\#\#\#} prefix from kept headers so they read as
prose. It also drops checkbox lines (\texttt{- [ ]} / \texttt{- [x]})
and collapses runs of blank lines. A real \cellB{}
example, verbatim from the released stripper:

\paragraph{Example (\texttt{django} PR 21426; 1{,}730 $\to$ 193
chars, $89\%$ reduction).} Raw (excerpt):
\begin{lstlisting}
Fixed #37130 -- Skipped DB cache deletion when culling offset is zero.

#### Trac ticket number
<!-- Replace XXXXX with the corresponding Trac ticket number. -->
<!-- Or delete the line and write "N/A - typo" for typo fixes. -->

ticket-37130

#### Branch description
Added a condition to bypass the deletion query when [...]

#### AI Assistance Disclosure (REQUIRED)
<!-- Please select exactly ONE of the following: -->
- [x] **No AI tools were used** in preparing this PR.
- [ ] **If AI tools were used**, I have disclosed which ones [...]

#### Checklist
- [x] This PR follows the [contribution guidelines](https://[...]
\end{lstlisting}
Stripped (complete):
\begin{lstlisting}
Fixed #37130 -- Skipped DB cache deletion when culling offset is zero.

Branch description
Added a condition to bypass the deletion query when the calculated culling offset evaluates to zero.
\end{lstlisting}

\section{The Behavior-Only Probe (\cellC{})}
\label{app:cellc}

\cellC{} holds $125$ queries ($87$ from \cellA{}, $38$ from \cellB{},
all ten repositories). For each, \texttt{gemma-4-31B-it} (temperature
$0$) rewrote the source query as a one- or two-sentence description of
the behavior, with every identifier, path, error string and ticket
reference removed. Gold files come from the source query. Because a
full description shrinks to a sentence or two, \cellC{} removes much
more than anchors. We use it only as one slice of the
decay analysis (\secref{law}) and as an anchor-removal test that also
loses other query content (\cref{app:genarm-moved}).

\section{Curated Multilingual Queries}
\label{app:curated-multiling}

Curated issue-symptom statements keep the dense premium but lose its
$\tau$-dependence: they share rare tokens with the target file as often
as raw PRs do, yet those tokens rarely anchor BM25. The $\tau$-effect is
therefore specific to lexically anchored queries.

\paragraph{Setup.} We score $134$ human-verified Multi-SWE-bench issue
statements on \texttt{vuejs/core}, \texttt{mui/material-ui} and
\texttt{tokio-rs/tokio} against the same corpora, with two encoders.

\paragraph{The premium stays; the slope disappears.} The premium is
$+0.03$ to $+0.18$, but its pooled slope on $\tau$ is flat ($+0.002$,
against $-0.164$ for the raw language slices).

\paragraph{Rare shared tokens no longer anchor BM25.} These queries have
the same $\tau$ as raw PRs. Yet three quarters return no \texttt{grep}
hit in the top ten, against $40\%$ for raw PRs, and BM25 scores far
lower (\tab{curated-multiling}).

Samples are small, query type is confounded with language, and
\texttt{tokio} does not show the gap. We read this as a limit on scope,
not a falsification.

\begin{table*}[t]
\centering
\small
\begin{sc}
\begin{adjustbox}{max width=\textwidth}
\begin{tabular}{llrrrr}
\toprule
Repo & Query type & $\tau$ & grep & BM25 & grep$=0$ \\
\midrule
\multicolumn{6}{l}{\emph{Same-repo comparison (identical corpus)}} \\
\texttt{vuejs/core} & raw (PR)        & 6.25 & 0.397 & 0.512 & 43\% \\
\texttt{vuejs/core} & curated (issue) & 6.13 & 0.104 & 0.130 & 84\% \\
\texttt{tokio}      & raw (PR)        & 6.37 & 0.386 & 0.569 & 34\% \\
\texttt{tokio}      & curated (issue) & 7.44 & 0.420 & 0.680 & 38\% \\
\midrule
\multicolumn{6}{l}{\emph{Pooled}} \\
raw (4 languages) & raw (PR)        & 7.05 & 0.325 & 0.453 & 40\% \\
Multi-SWE-bench   & curated (issue) & 6.93 & 0.137 & 0.250 & 76\% \\
\bottomrule
\end{tabular}
\end{adjustbox}
\end{sc}
\caption{Curated issue-symptom queries carry the same $\tau$ as raw PR
queries but retrieve far worse lexically. grep and BM25 are \ndcg{10};
grep$=0$ is the share of queries with no grep hit in the top ten;
\texttt{jina-code-0.5b}.}
\label{tab:curated-multiling}
\end{table*}

\section{Per-Repository Results and Example Queries}
\label{app:perrepo}

The premium is not driven by one repository: dense retrieval leads
BM25 at $c=1{,}000$ in every repository of \cellA{} but \texttt{requests},
and in all nine repositories of \cellB{} (\tab{perrepo}). Sizes range
from near zero to over $0.3$.

\begin{table*}[t]
\centering
\small
\begin{sc}
\begin{adjustbox}{max width=\textwidth}
\begin{tabular}{lrrrr}
\toprule
\textbf{Repo} & $n$ & \textbf{BM25} & \textbf{dense} & $\Delta$ \\
\midrule
\multicolumn{5}{l}{\emph{\cellA{} (curated, pre-cutoff)}} \\
\texttt{astropy}       &  21 & 0.490 & 0.563 & $+0.073$ \\
\texttt{django}        & 226 & 0.410 & 0.528 & $+0.119$ \\
\texttt{matplotlib}    &  34 & 0.378 & 0.593 & $+0.215$ \\
\texttt{pylint}        &   8 & 0.291 & 0.443 & $+0.152$ \\
\texttt{pytest}        &  19 & 0.293 & 0.423 & $+0.130$ \\
\texttt{requests}      &   8 & 0.543 & 0.534 & $-0.009$ \\
\texttt{scikit-learn}  &  18 & 0.561 & 0.586 & $+0.025$ \\
\texttt{sphinx}        &  44 & 0.304 & 0.489 & $+0.185$ \\
\texttt{sympy}         &  72 & 0.468 & 0.532 & $+0.064$ \\
\texttt{xarray}        &  22 & 0.404 & 0.615 & $+0.210$ \\
\midrule
\multicolumn{5}{l}{\emph{\cellB{} (raw PRs, post-cutoff)}} \\
\texttt{astropy}       &  14 & 0.405 & 0.495 & $+0.089$ \\
\texttt{django}        &  45 & 0.357 & 0.572 & $+0.215$ \\
\texttt{matplotlib}    &  13 & 0.329 & 0.690 & $+0.361$ \\
\texttt{pylint}        &  32 & 0.450 & 0.642 & $+0.192$ \\
\texttt{pytest}        &  34 & 0.586 & 0.634 & $+0.048$ \\
\texttt{requests}      &  27 & 0.374 & 0.579 & $+0.205$ \\
\texttt{scikit-learn}  &  36 & 0.348 & 0.595 & $+0.247$ \\
\texttt{sympy}         &  57 & 0.575 & 0.588 & $+0.013$ \\
\texttt{xarray}        &  30 & 0.416 & 0.592 & $+0.176$ \\
\bottomrule
\end{tabular}
\end{adjustbox}
\end{sc}
\caption{Per-repository \ndcg{10} for BM25, dense retrieval and the
premium (\texttt{jina-code-0.5b}, $c=1{,}000$, source-file labels). Dense
retrieval leads in every repository but one curated \texttt{requests}
slice of eight queries.}
\label{tab:perrepo}
\end{table*}

\subsection{Exemplar Queries of \cref{fig:decay}}
\label{app:intro_exemplars}

The dotted verticals in \cref{fig:decay} mark three real queries,
from rare-token to common-token. (A), $\tau{=}9.8$: ``simplify\_regexp
doesn't replace trailing groups'', BM25 $1.00$ vs.\ dense $0.33$.
(B), $\tau{=}8.5$: ``index refactor: more \_coord\_names
than \_variables on Dataset'', BM25 $0.61$ vs.\ dense $0.83$.
(C), $\tau{=}4.8$: ``Skipped DB cache deletion when
culling offset is zero'' (raw PR), BM25 $0.00$ vs.\ dense $1.00$.
The exemplars are extreme by construction; the claim of
\cref{sec:law} is about the mean curve.

\section{Premium Decay across Slices (Exploratory)}
\label{sec:law}
\label{app:law-slices}

Across $38$ evaluation slices the premium falls with $\tau$ in all but
one, and one common slope fits them. Because the predictor and its
normalization were chosen on these slices, this is exploratory; a
preregistered test on held-out repositories had the predicted sign but
missed its criterion.

\paragraph{Model and slices.} We fit the premium as one common function
of $\tau$'s within-repository percentile (IDF is not comparable across
corpora) plus a per-slice offset. The slices cover the five Python
mining regimes, all thirteen encoders on \cellA{} and \cellB{}, the
held-out pool and four other languages ($9{,}339$ query--slice
observations).

\begin{figure*}[t]
\centering
\includegraphics[width=0.98\textwidth]{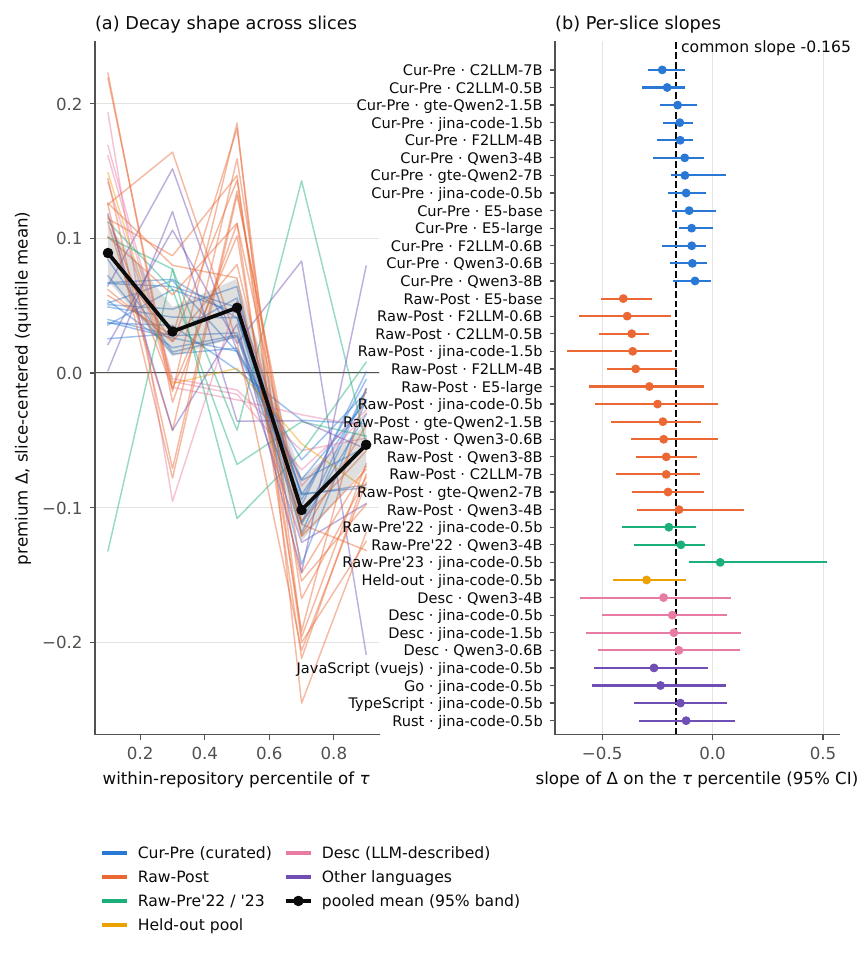}
\caption{The premium-decay regularity (exploratory). \emph{(a)}
Quintile means of $\Delta$ against the within-repository $\tau$
percentile for the 38 evaluation slices (thin lines, colored by slice
family), each centered on its own mean; black: pooled mean with its 95\%
band. \emph{(b)} Per-slice
slopes with bootstrap intervals; 37 of 38 are negative, and the dashed
line marks the common slope ($-0.165$). Source-file labels.}
\label{fig:law}
\end{figure*}

\paragraph{One slope fits all slices.} The per-slice slope is negative
in $37$ of $38$ slices. Encoders sharing a pool are not independent, so
the fairer count is nine of ten distinct query pools; the exception,
\cellRA{}, is small and near zero. The common slope is $-0.165$
\ndcg{10} across the full percentile range; intervals resample
repositories, then tasks, so encoders sharing a task move together.
Dropping any one repository barely changes the slope. A homogeneity
test does not reject a single slope. Per-slice slopes are noisy, so we
make the claim for the pooled data only (\fig{law}).

\paragraph{The preregistered held-out test was not met.} We applied the
frozen predictor to the $29$ repositories of the matched replication
(\secref{scaleup}) that no slice used. The slope is negative in all
seven sub-slices ($-0.109$ pooled), but its interval includes zero.
With about ten tasks per repository, the within-repository percentile is
too coarse to resolve a slope of this size.

\section{A Matched Replication at Scale}
\label{sec:scaleup}

Holding the change fixed, raw PR wording raises the premium over a
curated rewrite of the same change, for two encoders from different
families. A date effect of similar size cannot be excluded, and the
curated premium itself depends on the encoder.

\paragraph{Design.} We mine $360$ PRs from $36$ repositories in Go,
Java, Python, Rust and TypeScript/JavaScript, five before and five
after the encoders' cutoffs per repository; $351$ keep an indexed,
non-test gold file and are scored. Each task is queried twice,
against one gold set and one corpus at its base commit: once with the
raw PR text and once with a curated rewrite. \texttt{qwen35-122B} wrote
the rewrite from the PR text without seeing gold paths or patches, and
a separate call to the same model checked its fidelity. The curated
view is thus an LLM rewrite, not human curation. A repository-disjoint
development/test split was frozen before analysis.

\paragraph{Wording moves the premium within a task.} Raw wording raises
the premium for both \texttt{jina-code-0.5b} and
\texttt{Qwen3-Embedding-0.6B}, in about $70\%$ of repositories and on
the held-out test split (\tab{scaleup}). Both views share one gold set,
so the label rule cannot create this contrast. This is our strongest
evidence that wording moves the premium; the pooled ten-repository
comparison no longer separates curated from raw pools once labels are
consistent (\cref{app:q2-ext}).

\paragraph{Only the raw premium survives a change of encoder.} The
canonical encoder keeps a premium on both views, but the second
encoder's curated premium is indistinguishable from zero, as it is on
never-mined DeepSWE descriptions (\secref{external}). This is what we
expect if the premium is a property of the query.

\paragraph{Date is not separated from wording.} The post-minus-pre
contrast points the way contamination predicts and is about half the
size of the wording effect. Its interval includes zero, and a direct
comparison of the two sizes does not separate them. By language,
all ten estimates point the same way, but only Python (and, for the
second encoder, TypeScript/JavaScript and Rust) excludes zero on its
own. We therefore do not claim a five-language replication.

\begin{table*}[t]
\centering
\small
\begin{sc}
\begin{adjustbox}{max width=\textwidth}
\begin{tabular}{lcc}
\toprule
 & \textbf{jina-code-0.5b} & \textbf{Qwen3-Emb-0.6B} \\
\midrule
Wording effect     & $+0.091$ [$+0.024$, $+0.163$] & $+0.111$ [$+0.049$, $+0.176$] \\
\quad test split           & $+0.090$ [$+0.011$, $+0.172$] & $+0.119$ [$+0.043$, $+0.201$] \\
\quad repos positive       & $26/36$ & $25/36$ \\
\midrule
premium, curated view      & $+0.088$ [$+0.029$, $+0.145$] & $-0.015$ [$-0.077$, $+0.048$] \\
premium, raw view          & $+0.178$ [$+0.109$, $+0.245$] & $+0.096$ [$+0.026$, $+0.166$] \\
\midrule
date effect, curated       & $-0.044$ [$-0.147$, $+0.058$] & $-0.048$ [$-0.156$, $+0.055$] \\
date effect, raw           & $-0.028$ [$-0.140$, $+0.081$] & $-0.048$ [$-0.168$, $+0.068$] \\
$|$wording$|-|$date$|$, curated & $+0.046$ [$-0.075$, $+0.127$] & $+0.063$ [$-0.064$, $+0.143$] \\
\bottomrule
\end{tabular}
\end{adjustbox}
\end{sc}
\caption{Matched scale-up ($351$ of the $360$ tasks; the other nine have
no indexed, non-test gold file; $36$ repositories, five languages;
\ndcg{10}, source-file labels). Brackets: hierarchical
bootstrap 95\% CIs resampling repositories, then tasks. The date effect
is the within-repository post-minus-pre contrast.}
\label{tab:scaleup}
\end{table*}

\section{When BM25 Wins}
\label{sec:lexical-rescue}

BM25 wins a sizable minority of queries, but nothing we tried could
tell which ones from the query, so none of the predictors we tested
could use this headroom.

\paragraph{The headroom is real.} BM25 strictly beats dense retrieval on
about a quarter of \cellA{} and a fifth of \cellB{} queries, and picking
the better list per query would add about $0.09$ \ndcg{10} over dense
alone. The headroom is in the ranking order, not in coverage: the union
of both top-10 lists finds few gold files that dense retrieval misses.

\paragraph{The winner cannot be read from the query.} We tried
classifiers on seven structural query features (with or without
$\widehat\tau$ and length), a classifier on the query's embedding, and two
zero-shot LLM judges. Under leave-one-repository-out evaluation, none
predicts BM25 wins better than chance. Tuned on the
training repositories, each falls back to (nearly) always choosing
dense (\tab{router-pred}). Untuned, the judges pick
keyword search for most queries and lose over $0.1$ \ndcg{10}. Even $\tau$,
which uses the gold file, separates BM25 wins only weakly. The \cellB{} sample is
small, so a modest signal on raw queries cannot be excluded.

\begin{table}[t]
\centering
\small
\setlength{\tabcolsep}{3pt}
\begin{sc}
\begin{adjustbox}{max width=\columnwidth}
\begin{tabular}{lrr}
\toprule
\textbf{Predictor} & \textbf{AUROC} & \textbf{routed $-$ dense} \\
\midrule
Query features (GBT) & $0.51$ & $0.000$ \\
Query embedding (logistic) & $0.52$ & $-0.004$ \\
\texttt{qwen35-122B} judge & $0.54$ & $0.000$ \\
\texttt{gemma-4-31B} judge & $0.51$ & $-0.002$ \\
\bottomrule
\end{tabular}
\end{adjustbox}
\end{sc}
\caption{Predicting per query whether BM25 beats dense retrieval
(\cellA{}$+$\cellB{}, $604$ queries, leave one repository out). Pooled
AUROC on non-tied queries; every interval covers $0.5$. ``Routed $-$
dense'': \ndcg{10} change from switching to BM25 above a threshold tuned
on the training repositories. Query features: seven structural features
plus $\widehat\tau$, an identifier flag and length; embedding:
\texttt{jina-code-0.5b}. Other feature sets and models score
$0.47$--$0.51$.}
\label{tab:router-pred}
\end{table}


\section{Agent-Driven End-to-End Evaluation}
\label{sec:agent}
\label{app:agent}

Retrieval quality does not decide agent localization. For a SWE-agent
on the held-out pool, the choice of LLM moves exact-match far more than
the choice of search tool. Our first version of this evaluation was
invalid because the merged fix was reachable through \texttt{git}; all
numbers here come from the clean rerun.

\paragraph{Setup.} We draw stratified samples from the held-out pool
(\secref{heldout}), balanced across its nine repositories: $30$
queries for the model comparison and a nesting draw of $96$ for the
search-tool comparison. The agent is \texttt{mini-swe-agent}
\citep{minisweagent2025} in its raw-shell configuration (temperature
$0$, $30$ steps, $900$\,s per query). It sees the raw problem statement
and is scored on the files its final diff touches against the gold
files: precision, recall, any-hit (at least one gold file) and
exact-match (the gold set exactly).

\paragraph{The model matters most.} Across four open-weight
configurations (\tab{agent}), exact-match ranges from $0.033$
(\texttt{gemma-4-31B-it}) to $0.300$ (\texttt{kimi-k3}). Turning off
\texttt{qwen35-122B}'s reasoning phase changes exact-match by an amount
we cannot resolve, so the data do not show that thinking helps. The
spread across families mixes family with scale: it shows that the
served model matters, but not why. Low-recall trajectories share
one failure: the agent commits to a wrong file early and keeps editing
it.

\begin{table*}[t]
\centering
\small
\begin{sc}
\begin{adjustbox}{max width=\textwidth}
\begin{tabular}{lcccc}
\toprule
\textbf{Metric} & \textbf{qwen35-122B (think)} & \textbf{qwen35-122B (no-think)} & \textbf{kimi-k3 (think)} & \textbf{gemma-4-31B-it} \\
\midrule
mean precision   & $0.613$ & $0.561$ & $0.600$ & $0.350$ \\
mean recall      & $0.353$ & $0.266$ & $0.510$ & $0.468$ \\
any-hit rate     & $0.633$ & $0.600$ & $0.633$ & $0.733$ \\
exact-match rate & $0.133$ & $0.100$ & $0.300$ & $0.033$ \\
\bottomrule
\end{tabular}
\end{adjustbox}
\end{sc}
\caption{Model comparison on the $30$-query sample of the held-out pool
(clean rerun; fresh single-commit environments, no network). The
no-think column runs the same weights with the reasoning phase
disabled; \texttt{kimi-k3} replaces the withdrawn \texttt{kimi-k2.6} of
the superseded run and is reported in thinking mode only.}
\label{tab:agent}
\end{table*}

\subsection{Search Tools at a Fixed LLM}
\label{sec:agent-wrapper}

At a fixed LLM the search tool barely matters: four very different
tools land within about $0.04$ exact-match of each other, and no difference
between them is resolved.

\paragraph{Design.} We fix the LLM (\texttt{qwen35-122B}, thinking on)
and replace the shell's search with one tool, \texttt{repo\_search}.
Its backend is BM25, the canonical dense encoder, or a router that uses
$\widehat\tau$ to pick one of the two per query. The restriction is
enforced: \texttt{find}, \texttt{grep}, \texttt{rg} and \texttt{git
grep} are shadowed on \texttt{PATH}, and attempts around them are
logged.

\paragraph{No tool wins.} The unrestricted shell scores lowest in point
estimate, but no paired difference excludes zero. The ordering of the
three tools also changes between the $30$- and $96$-query samples
(\tab{agent-wrapper}). At $30$ steps, what limits the agent is which
file the LLM commits to, not how it found candidates. This does not
show that agents write their own anchors; that argument rests on
\cref{sec:genarm} and \cref{sec:external}.

\begin{table*}[t]
\centering
\small
\begin{sc}
\begin{adjustbox}{max width=\textwidth}
\begin{tabular}{lcccc}
\toprule
\textbf{Metric} & \textbf{raw shell} & \textbf{BM25 index} & \textbf{dense} & \textbf{router} \\
\midrule
mean precision   & $0.517$ & $0.475$ & $0.519$ & $0.541$ \\
mean recall      & $0.359$ & $0.385$ & $0.370$ & $0.392$ \\
any-hit rate     & $0.573$ & $0.604$ & $0.615$ & $0.625$ \\
exact-match rate & $0.135$ & $0.167$ & $0.146$ & $0.177$ \\
\bottomrule
\end{tabular}
\end{adjustbox}
\end{sc}
\caption{Search-tool comparison at a fixed LLM (\texttt{qwen35-122B},
thinking on), $96$ queries, clean rerun. Only the search affordance
varies; the tool restriction is enforced on \texttt{PATH}.}
\label{tab:agent-wrapper}
\end{table*}

\paragraph{The $\tau$-effect reaches the agent on any-hit only.} A
preregistered test predicted that the dense tool beats the BM25 tool on
queries with few rare anchors (bottom $\widehat\tau$ quartile) and not
on the top quartile. On any-hit it holds: dense is well ahead at the
bottom and behind at the top, so the flat pooled comparison hides a
large swing. On exact-match the contrast is unresolved: finding some
gold file more often does not mean reproducing the whole gold set.

\subsection{Task-Environment Contamination}
\label{sec:agent-contamination}

Our first version of this evaluation leaked the answer, and the leak
reversed its main result. We report it because the mistake is easy to
make in any benchmark built from merged changes.

\paragraph{What went wrong.} Tasks were materialized with
\texttt{git worktree add} at the base commit. This pins the files but
shares the object store, so the merged fix stayed reachable through
\texttt{git log -{}-all} or \texttt{git show}. In $8$ of $30$ shell
trajectories the agent read a later commit covering the whole gold set.
The tool restriction was also only a request in the prompt, which the
agent ignored in most trajectories.

\paragraph{The fix, and what it changed.} Each task is now exported with
\texttt{git archive} into a fresh repository with one synthetic commit
and no remotes, checked per query. The tool restriction is enforced,
and the shell has no network access; this also closes the route by
which one query had fetched its own pull-request diff. Raw-shell exact-match
fell from $0.333$ to $0.133$, while the BM25 tool, which had nothing to
gain from the leak, stayed at $0.167$. The apparent advantage of the
shell was therefore the leak.

\paragraph{Limitations.} The model comparison uses $30$ queries, so only
its largest differences resolve. Scoring is file-level: any-hit is a
lenient ceiling and no execution-based check is run. The search tools
return $20$ chunk hits that are not deduplicated to files.


\section{Never-Mined Queries (DeepSWE)}
\label{sec:external}

DeepSWE task descriptions are written to specify work, not mined from a
completed change. On them the premium does not transfer and BM25 wins
far more often, which fits the view that the premium depends on how
queries are written.

\paragraph{Setup.} $109$ of DeepSWE's $113$ tasks pass our qrel audit
($88$ repositories, five languages: Python, JavaScript, TypeScript, Go
and Rust). The gold files are those of one
reference patch, so we measure reference-edit localization rather than
exhaustive relevance. The retrieval harness is that of \secref{methods}.

\paragraph{The premium does not transfer.} The canonical encoder's
premium, $+0.121$ and $+0.152$ on our curated and raw pools, is not
distinguishable from zero here, and \texttt{Qwen3-Embedding-0.6B} falls
below BM25 (\tab{external}). This is the sharpest limit on the premium
we found, and it matches the within-task wording effect of
\secref{scaleup}.

\paragraph{Lexical rescue is a property of the query.} BM25 strictly
wins $38$--$52\%$ of queries, against about a fifth to a quarter on our
pools, and the two encoders' BM25-win sets largely coincide.

\paragraph{Corpus size does not explain it.} If anything the premium
shrinks as corpora grow, though no size tercile resolves.

\begin{table*}[t]
\centering
\small
\begin{sc}
\begin{adjustbox}{max width=\textwidth}
\begin{tabular}{lccccc}
\toprule
\textbf{Encoder} & \textbf{dense} & \textbf{BM25} & \textbf{grep} & \textbf{dense$-$BM25} & \textbf{BM25 wins} \\
\midrule
jina-code-0.5b   & $0.481$ & $0.445$ & $0.435$ & $+0.036$ [$-0.030$, $+0.100$] & $37.6\%$ \\
Qwen3-Emb-0.6B   & $0.367$ & $0.445$ & $0.435$ & $-0.078$ [$-0.144$, $-0.013$] & $52.3\%$ \\
\bottomrule
\end{tabular}
\end{adjustbox}
\end{sc}
\caption{Retrieval on DeepSWE's original task descriptions (\ndcg{10};
$68$ tasks in $56$ repositories whose corpus is identical across the two
encoder runs, so the lexical columns are shared). Brackets:
repository-clustered paired bootstrap 95\% CIs.}
\label{tab:external}
\end{table*}

\end{document}